\documentclass[12pt]{article}
\usepackage{epsfig}
\usepackage{amsmath}
\usepackage{amssymb}
\usepackage{graphicx}
\usepackage{lscape}
\usepackage {subfigure}
\usepackage{hyperref, cite}
\def\be{\begin{equation}}
\def\ee{\end{equation}}
\def\ba{\begin{array}}
\def\ea{\end{array}}
\def\beqn{\begin{eqnarray}}
\def\eeqn{\end{eqnarray}}

\def\bt{\begin{tabular}}
\def\et{\end{tabular}}
\def\bc{\begin{center}}
\def\ec{\end{center}}
\usepackage{epstopdf}
\begin{document}
\title{Rephasing invariant quartets  in texture zero neutrino mass matrix and CP violation }
\author{Madan Singh$^{*} $\\
\it Department of Physics, M. N. S Government College Bhiwani, Bhiwani,\\
\it Haryana,127021, India.\\
\it $^{*}$singhmadan179@gmail.com
}
\maketitle
\begin{abstract}
In the quark sector, Jarlskog rephasing invariant  $\rm{J_{CP}}$ has important implications for the CP violation as well as phase structure of the
quark mass matrices. In fact all CP violating effects in this sector are proportional to the magnitude of the imaginary part of $\rm{J_{CP}}$. 
  With the observation of non-zero and relatively large reactor mixing angle $\theta_{13}$,  it is widely believed that CP might be violated in the lepton sector. The recent global best fits of neutrino oscillation indicates the preference of  CP violation  over CP conservation at relatively large confidence level.  These experimental observations warrant  to re-investigate the CP-odd weak basis invariants  at low energies, and also explore their implications  for CP violation in the context of  texture two zero neutrino mass matrix.  
\end{abstract}

\section{Intoduction}
CP violation offers a powerful tool to explore the flavor sector of the Standard Model (SM) \cite{1, 2, 3, 4, 5}, and to search the signals of New Physics (NP). After a long and exciting history of K-decay study, the stage is now governed by the decays of B mesons. The efforts at the $e^{+} e^{-}$ B factories with their detectors BaBar (SLAC) and Belle(KEK),CP violation is now well established in the B meson system.
      In the Standard Model, at present the only way to understand CP violation seems the Cabibbo-Kobayashi-Maskawa (CKM) \cite{6} mechanism. This CKM picture of CP violation is supported by the precision measurements of sin2$\beta$ from the CP asymmetry in the decay B$\rightarrow \psi$ K. This picture is also supported by reconstruction of the unitarity triangle through global fits by various well known groups like Particle Data Group (PDG) \cite{7}, CKMfitter \cite{8}, UTfit \cite{9}  and HFAG \cite{10}. In the  CKM mechanism, phase $\delta$ is a only source of CP violation in the case of quarks, and is defined through the Cabibbo-Kobayashi-Maskawa (CKM) mixing matrix, $\rm{V_{CKM}}$ given by \cite{6}.  In the lepton sector, standard model does not allow the CP violation. However, in the spirit of quark-lepton symmetry, it is natural to expect an entirely analogous mechanism to arise in the lepton sector, leading to leptonic CP violation.
       
 The  experimental observation of neutrino oscillation,  \cite{11, 12, 13, 14}, in this regard,  not only suggests that neutrinos are massive,  but also open up  the possibility of CP violation in leptonic sector.   This has further triggered a sudden spurt of activity on the theoretical as well as experimental front to identify the possibilities beyond the standard model, where leptonic CP violation would be observed. In comparison to the case of hadrons, the parameterization of CP violation in the leptonic case is complicated by several factors. In particular, if the neutrinos are Majorana particles then in contrast to the case of CKM matrix, the corresponding leptonic Pontecorvo-Maki-Nakagawa-Sakata (PMNS) mixing matrix has two additional phases apart from an analogous CKM phase $\delta$. For example, the leptonic mixing matrix $\rm{V_{PMNS}}$ is given by \cite{15, 16}      

\begin{equation*}
\rm{V_{PMNS}= \left(
\begin{array}{ccc}
 c_{12}c_{13}& s_{12}c_{13}& s_{13} \\
-c_{12}s_{23}s_{13}-s_{12}c_{23}e^{-i\delta} & -s_{12}s_{23}s_{13}+c_{12}c_{23}e^{-i\delta} & s_{23}c_{13}\\
 -c_{12}c_{23}s_{13}+s_{12}s_{23}e^{-i\delta}& -s_{12}c_{23}s_{13}-c_{12}s_{23}e^{-i\delta}& c_{23}c_{13} \\
\end{array}
\right)               
\left(
\begin{array}{ccc}
e^{i\rho}  & 0& 0 \\
0 & e^{i\sigma} & 0\\
0& 0& 1 \\
\end{array}
 \right).}
\end{equation*}
Here, $c_{ij} = \cos \theta_{ij}$, $s_{ij}= \sin \theta_{ij}$ for $i, j=1, 2, 3$, and $\delta$, $\sigma$ denote the Dirac and Majorana type CP  phase, respectively. The main difference between a Majorana phase and Dirac phase is that the Majorana phases do not affect any lepton number conserving process like neutrino oscillations. On the other hand, the Dirac phases may contribute to both lepton number conserving as well as lepton number violating process. 
 
 The search for CP violation in the leptonic sector at low energies is at present one of the major experimental challenges in neutrino physics. Experiments with superbeams and neutrino beams from muon
storage rings (neutrino factories) have the potential  to measure directly the Dirac
phase $\delta$ through CP and T asymmetries or indirectly through neutrino oscillations. An alternative method
\cite{17} is to measure the area of unitarity triangles defined for the leptonic sector \cite{18}. In addition, the effects of Majorana type phases may arise in neutrinoless double beta decay ($0\nu \beta \beta$) processes. The observation of such processes would establish the Majorana nature of neutrinos and, possibly, provide some information on the Majorana CP phases. Thus, neutrino physics provides an invaluable tool for the investigation of leptonic CP violation at low energies apart from having profound implications for the physics of the early universe understand the baryon number asymmetry in the universe.\\

In the analysis of lepton flavor models, a useful approach while addressing the question of CP violation is the construction of the CP-odd weak basis (WB) invariants, which are independent of basis choice and phase convention. The invariants should vanish if CP is an exact symmetry of the theory. Thus, in CP violating theories which contain several phases, invariants constitute a powerful tool to probe whether a particular texture zero mass matrix leads to leptonic CP violation at high and/or low energies.  For texture  two zero  mass matrix , S Dev et. al \cite{19} have  presented the WB invariants in terms of elements of neutrino mass matrix in the basis, where charged lepton mass matrix is diagonal, and subsequently derived the CP invariance conditions for all the viable ansatz. Similar analysis has been carried out by U. Sarkar and S. Singh \cite{20} for texture two zero neutrino mass matrices. In addition, the implications of CP-odd invariant parameters have been presented for other textures as well \cite{21}. 

The non-zero and large "reactor mixing angle", $\theta_{13}$ \cite{22,23,24,25}, has strengthened our belief that CP might be violated in the leptonic sector.  The violation of  CP symmetry is also indicative  in
T2K, Superkomiokande and NO$\nu$A experiments as well as in the recent global fit analyses \cite{26,27}. 
Therefore, it becomes imperative to re-look at these CP–odd rephasing invariants for finding the CP violation in the context of texture two zero Majorana neutrino mass matrices.

\section{Weak basis invariants and CP violation}
At the low-energy scale, lepton masses, flavor mixing angles and CP-violating phases are governed
by the following effective Lagrangian \cite{28}
\begin{equation}\label{eq1}
\rm{-L_{mass}=- \sum_{l, l^{'}=e,\mu, \tau} \overline{l}_{ L} m_{l} l_{ R}+\frac{1}{2}\overline{\nu_{L}}m_{\nu} \nu_{L}^{C}+ \textit{h.c.},}
\end{equation}\\
where $\rm{\nu_{L}^{C}\equiv C\overline{\nu_{L}}^{T}}$ with C=i$\gamma^{2}\gamma^{0}$ being the charge-conjugation matrix. $m_{l}$ and $m_{\nu}$ stand for the
charged-lepton mass matrix and the Majorana neutrino mass matrix, respectively.
In the  basis, where $m_{l}$  is diagonal, one can write the complex symmetric Majorana neutrino mass matrix as
\begin{small}
\begin{equation}\label{eq2}
\rm{m_{\nu}\equiv\left(
\begin{array}{ccc}
   m_{ee}& m_{e \mu}& m_{e \tau} \\
   & m_{\mu \mu} & m_{\mu \tau}\\
 & & m_{\tau \tau} \\
  \end{array}
  \right)=V_{PMNS} \left(\begin{array}{ccc}
   m_{1}&0& 0 \\
   & m_{2} & 0\\
 & & m_{3} \\
  \end{array}
  \right)V_{PMNS}^{T},}
  \end{equation}
  \end{small},
  where, $\rm{m_{1}, m_{2}}$ and $\rm{m_{3}}$ are neutrino masses. For the texture two zero ansatz, neutrinos masses can be calculated, considering  the relations provided in Ref.\cite{29}.\\
From Eq.(\ref{eq2}) , it is apparent that $\rm{m_{\nu}}$ depends on seven low energy physical parameters: three neutrino masses ($\rm{m_{1},m_{2}, m_{3}})$, three mixing angles ($ \theta_{12},\theta_{23}, \theta_{13})$,  three CP violating phases ($\delta, \rho, \sigma$), therefore, one can trivially derive each element of $\rm{m_{\nu}}$ in terms of these parameters, 
\begin{small}
\begin{eqnarray*}
&&\rm{m_{ee}=m_{1}c_{12}^{2}c_{13}^{2}e^{2i\rho}+m_{2}s_{12}^{2}c_{13}^{2}e^{2i\sigma}+m_{3}s_{13}^{2}},\\
&&\rm{m_{e\mu}=m_{1}(-c_{13}s_{13}c_{12}^{2}s_{23}e^{2i\rho}-c_{13}c_{12}s_{12}c_{23}e^{i(2\rho-\delta)})+m_{2}(-c_{13}s_{13}s_{12}^{2}s_{23}e^{2i\sigma}+c_{13}c_{12}s_{12}c_{23}e^{i(2\sigma-\delta)})}
\nonumber  \\ 
 &&~~~~~~~~
\rm{+m_{3}c_{13}s_{13}s_{23}},\\
&&\rm{m_{\mu\mu}=m_{1}(c_{12}s_{13}s_{23}e^{i\rho}+c_{23}s_{12}e^{i(\rho-\delta)})^{2}+ m_{1}(s_{12}s_{13}s_{23}e^{i\sigma}-c_{23}c_{12}e^{i(\sigma-\delta)})^{2}}
\nonumber  \\ 
 &&~~~~~~~~
\rm{+m_{3}c_{13}^{2}s_{23}^{2}},\\
&&\rm{m_{\mu\tau}=m_{1}(c_{12}^{2}c_{23}s_{23}s_{13}^{2}e^{2i\rho}+s_{13}c_{12}s_{12}(c_{23}^{2}-s_{23}^{2})e^{i(2\rho-\delta)}-c_{23}s_{23}s_{12}^{2}e^{2i(\rho-\delta)})+}
\nonumber  \\ 
 &&~~~~~~~~
\rm{m_{2}(s_{12}^{2}c_{23}s_{23}s_{13}^{2}e^{2i\sigma}-s_{13}c_{12}s_{12}(c_{23}^{2}-s_{23}^{2})e^{i(2\sigma-\delta)}-c_{23}s_{23}c_{12}^{2}e^{2i(\sigma-\delta)})}
\nonumber  \\ 
 &&~~~~~~~~
\rm{+m_{3}s_{23}c_{23}c_{13}^{2})},
\end{eqnarray*}
\end{small}
where, $\rm{m_{\mu\mu}}$ $(\rm{m_{e\mu}})$ and $\rm{m_{\tau\tau}}$ $(\rm{m_{e\tau}})$ are related through $\mu-\tau$ permutation symmetry. The application of permutation symmetry,  lead to the following relations among the neutrino oscillation parameters 
\begin{equation}\label{eq3}
\rm{\theta_{12}^{X}=\theta_{12}^{Y},}\;
\rm{\theta_{23}^{X}=90^{0}-\theta_{23}^{Y},}\;
\rm{\theta_{13}^{X}=\theta_{13}^{Y}, \delta^{X}=\delta^{Y}+ 180^{0},}
\end{equation}
where, X and Y superscripts denote the elements (or ansatz) related by  permutation symmetry. On employing Eq.(\ref{eq3}), the expressions  for   $\rm{m_{\tau\tau}}$ $(\rm{m_{e\tau}})$ can  trivially be obtained from $\rm{m_{\mu\mu}}$ $(\rm{m_{e\mu}})$.

 In the following discussion, I shall define the  WB invariants \cite{30,31, 32, 33}, which must be non-zero for detecting CP violation in lepton sector  at low energy. The relevance of CP odd WB invariants in the analysis of the texture zero ansatze is due to the fact that texture zeros lead to a decrease in the number of the independent CP violating phases. The number of CP odd invariants coincides with the number of CP-violating phases which arise in the lepton mixing of charged  weak current, after all lepton masses have been diagonalised. In Ref. \cite{34}, B. Yu and S. Zhou prove through the numerical analysis that the minimal set of CP-odd invariants , which lead to the CP conservation, are three, and hence the number is not accidental. 
 In pursuit for CP violation , it is imperative to discuss these invariants in terms of neutrino mass matrix in the weak basis. 
 
The invariant, which  is sensitive to Dirac phase only, is given as
\begin{equation}\label{eq4}
\rm{I_{1}=Tr[m_{\nu } m_{\nu }^{\dagger}, m_{l}m_{l}^{\dagger}]^{3}.}
\end{equation}
The invariant $\rm{I_{1}}$ plays a role entirely analogous to the one played \cite{30} by $\rm{Tr[m_{d} m_{d }^{\dagger}, m_{u}m_{u}^{\dagger}]^{3}}$ in the quark sector with three generations.  The computation of CP violation through $\rm{I_{1}}$ is possible only in the "weak basis", which requires $m_{l}$ to be real and diagonal.
In the weak basis, Eq.(\ref{eq4}) can be deduced  as 
\begin{equation*}
\rm{I_{1}=-6i\Delta_{e\mu}\Delta_{\mu\tau}\Delta_{\tau e}Im(h_{12}h_{23}h_{31})},
\end{equation*}
\begin{equation}\label{eq5}
\rm{\qquad= -6i\Delta_{e\mu}\Delta_{\mu\tau}\Delta_{\tau e} |h_{12}||h_{23}||h_{31}|sin\phi}.
\end{equation} 
where, $\rm{\Delta_{ij}=m_{j}^{2}-m_{i}^{2}}$, ij run over the pairs, $e\mu,\mu\tau, \tau e$, and 
$\rm{\phi=arg(h_{12}h_{23}h_{31})}$ is a physical phase responsible for CP violation.\\
In Eq. (\ref{eq5}), $ \rm{h \equiv m_{\nu}m_{\nu}^{\dagger}}$, is a hermitian matrix , and its elements are given as
\begin{eqnarray*}
\rm{h_{11}=|m_{ee}|^{2}+|m_{e\mu}|^{2}+|m_{e\tau}|^{2}},\\
\rm{h_{12}=m_{ee} m_{e\mu}^{*}+m_{e\mu} m_{\mu\mu}^{*}+m_{e\tau} m_{\mu \tau}^{*}},\\
\rm{h_{13}=m_{ee} m_{e\tau}^{*}+m_{e\mu} m_{\mu\tau}^{*}+m_{e\tau} m_{\tau \tau}^{*},}\\
\rm{h_{21}=m_{e\mu} m_{ee}^{*}+m_{\mu\mu} m_{e\mu}^{*}+m_{\mu\tau} m_{e \tau}^{*},}\\
\rm{h_{22}=|m_{e\mu}|^{2}+|m_{\mu \mu}|^{2}+|m_{\mu\tau}|^{2},}\\
\rm{h_{23}=m_{e\mu} m_{e\tau}^{*}+m_{\mu\mu} m_{\mu\tau}^{*}+m_{\mu\tau} m_{\tau \tau}^{*},}\\
\rm{h_{31}=m_{e\tau} m_{ee}^{*}+m_{\mu\tau} m_{e\mu}^{*}+m_{\tau\tau} m_{e \tau}^{*},}\\
\rm{h_{32}=m_{e\tau} m_{e\mu}^{*}+m_{\mu\tau} m_{\mu\mu}^{*}+m_{\tau\tau} m_{\mu \tau}^{*},}\\
\rm{h_{33}=|m_{e\tau}|^{2}+|m_{\mu \tau}|^{2}+|m_{\tau\tau}|^{2}}.\\
\end{eqnarray*}
From the above, it is explicit that the diagonal elements are real,  implying  that $\rm{I_{1}}$ does not depend on the Majorana phases $\rho$ and $\sigma$ appearing in the leptonic mixing matrix.  The  invariant is sensitive to the Dirac type CP phase $\delta$ and vanishes for $\delta$ = 0. This can be shown  through the relationship between  $\rm{I_{1}}$ and  Jarlskog rephasing invariant, $\rm{J_{CP}}$ for lepton sector. \\
 By definition,  $\rm{J_{CP}}$ can be expressed, in terms of $\delta$  as
\begin{equation}\label{eq6}
\rm{J_{CP}= \frac{1}{8}s_{2(12)}s_{2(23)}s_{2(13)}c_{13} \sin\delta},
\end{equation}
where, $s_{2(ij)}\equiv sin2\theta_{ij}$, $i,j=1,2,3$.\\
 Also $J_{CP}$ is an 'invariant function' of mass matrices, and is related to the mass matrices as
\begin{eqnarray}\label{eq7}
&&
\rm{det C=-2J_{CP} \Delta_{e\mu}\Delta_{\mu\tau}\Delta_{\tau e}  \Delta_{12} \Delta_{23}\Delta_{31}},
\end{eqnarray} 
where, $\rm{C \equiv i[m_{\nu}m_{\nu}^{\dagger}, m_{\l}m_{l}^{\dagger}]}$, and,
$\Delta_{12}, etc$ are analogue to the $\Delta_{ij}$ as defined earlier. 
The commutator C, is by definition, hermitian and traceless. Thus eigen values are real. In fact they are measurable, even though C itself is not a measurable. The determination of any traceless $3\times 3$ may be computed from trace of the third power of the matrix, i.e., $ \rm{detC=\frac{1}{3}Tr(C^{3})}$, therefore I have
$ \rm{det C=-\frac{i}{3}Tr[m_{\nu}m_{\nu}^{\dagger}, m_{\l}m_{l}^{\dagger}]^{3}}$,
which is valid for any traceless 3 $\times$ 3 matrix.
Therefore from Eqs.(\ref{eq7}), I get
\begin{equation}\label{eq8}
\rm{I_{1}= -6iJ_{CP} \Delta_{e\mu}\Delta_{\mu\tau}\Delta_{\tau e}  \Delta_{12} \Delta_{23}\Delta_{31}}.
\end{equation}
The above relation shows  that $\rm{I_{1}}$ is directly proportional to $\delta$.\\
 Using Eqs. (\ref{eq5}), \ref{eq8}, I arrive at
\begin{eqnarray}\label{eq9}
&&\rm{J_{CP}=\frac{Im(h_{12}h_{23}h_{31})}{\Delta_{sol} \Delta_{atm}^{2}}},
\nonumber  \\ 
 &&~~~~~
 =\rm{ \frac{|h_{12}||h_{23}||h_{31}|sin\phi}{\Delta_{sol} \Delta_{atm}^{2}}},
\end{eqnarray}
where, $\Delta_{sol}\equiv \Delta_{12}$ and $\Delta_{atm} \equiv \Delta_{23}\simeq \Delta_{31} $, are solar and atmospheric neutrino mass squared differences.
\\
 From Eq.(\ref{eq9}), it is clear that $\rm{J_{CP}}$ can be evaluated directly, however the computation of $\delta$ is  not straightforward. For that purpose, one requires an expansion of $\rm{I_{1}}$ in terms of $\delta$ using Eq.(\ref{eq5}).  The discussion, in this regard, is contained in section III.\\ 

The other two invariants, $\rm{I_{2}}$ and $\rm{I_{3}}$, are sensitive to both Dirac as well as Majorana type CP phases. The  invariant $\rm{I_{2}}$ is given as
\begin{equation}\label{eq10}
\rm{I_{2}=Im( Tr[m_{l } m_{l}^{\dagger} m_{\nu}^{*}m_{\nu}m_{\nu}^{*}m_{l}^{T}m_{l}^{*}m_{\nu} ])}.
\end{equation}
The above invariant was computed, for the first time, to derive the necessary and sufficient condition for CP invariance in the framework of two  neutrinos[\cite{31}]. In this framework, CP violation  can occur only due to Majorana type phase. In the weak basis, one can re-write  $\rm{I_{2}}$ as a function  of the elements of $m_{\nu}$, 
\begin{eqnarray}\label{eq11}
&&
\rm{I_{2}= m_{e}^{4}m_{ee}H_{11}+m_{\mu}^{4}m_{\mu\mu}H_{22}+m_{\tau}^{4}m_{\tau \tau}H_{33}
 +m_{e}^{2}m_{\mu}^{2}m_{e\mu}(H_{12}+H_{21})}\nonumber  \\
&& \quad\quad \qquad \rm{+m_{\mu}^{2}m_{\tau}^{2}m_{\mu\tau}(H_{23}+H_{32})+m_{e}^{2}m_{\tau}^{2}m_{e\tau}(H_{13}+H_{31})},
\end{eqnarray} 
where, H is a complex matrix, and its elements are given as
\begin{equation*}
\rm{H_{11}=m_{ee}^{*}h_{11}+m_{e\mu}^{*}h_{21}+m_{e\tau}^{*}h_{31}},
\end{equation*}
\begin{equation*}
\rm{H_{12}=m_{ee}^{*}h_{12}+m_{e\mu}^{*}h_{22}+m_{e\tau}^{*}h_{32}},
\end{equation*}
\begin{equation*}
\rm{H_{13}=m_{ee}^{*}h_{13}+m_{e\mu}^{*}h_{23}+m_{e\tau}^{*}h_{33}},
\end{equation*}
\begin{equation*}
\rm{H_{21}=m_{e\mu}^{*}h_{11}+m_{\mu\mu}^{*}h_{21}+m_{\mu\tau}^{*}h_{31}},
\end{equation*}
\begin{equation*}
\rm{H_{22}=m_{e\mu}^{*}h_{12}+m_{\mu\mu}^{*}h_{22}+m_{\mu\tau}^{*}h_{32}},
\end{equation*}
\begin{equation*}
\rm{H_{23}=m_{e\mu}^{*}h_{13}+m_{\mu\mu}^{*}h_{23}+m_{\mu\tau}^{*}h_{33}},
\end{equation*}
\begin{equation*}
\rm{H_{31}=m_{e\tau}^{*}h_{11}+m_{\mu\tau}^{*}h_{21}+m_{\tau\tau}^{*}h_{31}},
\end{equation*}
\begin{equation*}
\rm{H_{32}=m_{e\tau}^{*}h_{12}+m_{\mu\tau}^{*}h_{22}+m_{\tau\tau}^{*}h_{32}},
\end{equation*}
\begin{equation*}
\rm{H_{33}=m_{e\tau}^{*}h_{13}+m_{\mu\tau}^{*}h_{23}+m_{\tau\tau}^{*}h_{33}}.
\end{equation*}
From the above, it is apparent that all the elements are complex. Therefore H depends on both Dirac as well as Majorana type CP  phases.\\
On substituting  $\rm{m_{\nu}m_{\nu}^{\dagger}}$ with $\rm{m_{\nu}(m_{l}m_{l}^{\dagger})^{*}m_{\nu}^{\dagger}}$ in Eq.(\ref{eq4}),  $\rm{I_{3}}$ can  be trivially expressed as, 
\begin{equation}\label{eq12}
\rm{I_{3}\equiv Tr[m_{\nu}m_{l}^{T}m_{l}^{*}m_{\nu}^{\dagger}, m_{l}m_{l}^{\dagger}]^{3}}.
\end{equation}
The computation  of  above invariant  predicts CP violation for  three or more generation of Majorana neutrinos even in the limit of complete neutrino mass degeneracy \cite{32}. This is  contrary to  the case of Dirac neutrinos, where in the limit of exact degeneracy it is well known that there is no CP violation or physical lepton mixing, for an arbitrary number of generations.\\ 
  In the weak basis, invariant $\rm{I_{3}}$ can be written as
\begin{equation*}
\rm{I_{3}=6i\Delta_{e\mu}\Delta_{\mu\tau}\Delta_{\tau e}Im(h_{12}^{'}h_{23}^{'}h_{31}^{'})},
\end{equation*}
\begin{equation}\label{eq13}
\qquad = \rm{6i\Delta_{e\mu}\Delta_{\mu\tau}\Delta_{\tau e}|h_{12}^{'}||h_{23}^{'}||h_{31}^{'}|sin\theta},
\end{equation}
where, $\rm{\theta=arg(h_{12}^{'}h_{23}^{'}h_{31}^{'}}$) is complex phase responsible for CP violation.\\
 Similar to $h$, $h^{'}$ is also hermitian matrix, 
 and its elements are given as 
\begin{equation*}
\rm{h_{11}^{'}=m_{e}^{2}|m_{ee}|^{2}+m_{\mu}^{2}|m_{e\mu}|^{2}+m_{\tau}^{2}|m_{e\tau}|^{2},}\\
\end{equation*}
\begin{equation*}
\rm{h_{12}^{'}=m_{e}^{2}m_{ee} m_{e\mu}^{*}+m_{\mu}^{2}m_{e\mu} m_{\mu\mu}^{*}+m_{\tau}^{2}m_{e\tau} m_{\mu \tau}^{*},}\\
\end{equation*}
\begin{equation*}
\rm{h_{13}^{'}=m_{e}^{2}m_{ee} m_{e\tau}^{*}+m_{\mu}^{2}m_{e\mu} m_{\mu\tau}^{*}+m_{\tau}^{2}m_{e\tau} m_{\tau \tau}^{*},}\\
\end{equation*}
\begin{equation*}
\rm{h_{21}^{'}=m_{e}^{2}m_{e\mu} m_{ee}^{*}+m_{\mu}^{2}m_{\mu\mu} m_{e\mu}^{*}+m_{\tau}^{2}m_{\mu\tau} m_{e \tau}^{*},}\\
\end{equation*}
\begin{equation*}
\rm{h_{22}^{'}=m_{e}^{2}|m_{e\mu}|^{2}+m_{\mu}^{2}|m_{\mu \mu}|^{2}+m_{\tau}^{2}|m_{\mu\tau}|^{2},}\\
\end{equation*}
\begin{equation*}
\rm{h_{23}^{'}=m_{e}^{2}m_{e\mu} m_{e\tau}^{*}+m_{\mu}^{2}m_{\mu\mu} m_{\mu\tau}^{*}+m_{\tau}^{2}m_{\mu\tau} m_{\tau \tau}^{*},}\\
\end{equation*}
\begin{equation*}
\rm{h_{31}^{'}=m_{e}^{2}m_{e\tau} m_{ee}^{*}+m_{\mu}^{2}m_{\mu\tau} m_{e\mu}^{*}+m_{\tau}^{2}m_{\tau\tau} m_{e \tau}^{*},}\\
\end{equation*}
\begin{equation*}
\rm{h_{32}^{'}=m_{e}^{2}m_{e\tau} m_{e\mu}^{*}+m_{\mu}^{2}m_{\mu\tau} m_{\mu\mu}^{*}+m_{\tau}^{2}m_{\tau\tau} m_{\mu \tau}^{*},}\\
\end{equation*}
\begin{equation*}
\rm{h_{33}^{'}=m_{e}^{2}|m_{e\tau}|^{2}+m_{\mu}^{2}|m_{\mu \tau}|^{2}+m_{\tau}^{2}|m_{\tau\tau}|^{2},}\\
\end{equation*}
where, $\rm{m_{e}, m_{\mu}}$ and $\rm{m_{\tau}}$ denote the  electron, muon and tau neutrino, respectively.\\ 
In Ref. \cite{19}, S. Dev have  remarked that there is only one complex phase which is responsible for CP violation in leptonic sector, while  Dirac and  Majorana type CP phases can not be separated. In this paper, I shall update the result of Ref. \cite{19}  to compute the CP phases. \\

\section{Texture two zero ansatze, WB invariants and CP violation}
To investigate flavour symmetries and relations that determine the CP violation in general (either through Dirac phase or through Majorana type phases), it is imperative to consider effective neutrino mass matrix $\rm{m_{\nu}}$. However, the argument of the element of $\rm{m_{\nu}}$, which defines the CP violating phases
donot have any physical meaning since they are not invariant under the rephasing transformation, 
\begin{small}
\begin{equation}\label{eq14}
\rm{m_{\nu}\rightarrow diag(e^{i\phi_{e}}, e^{i\phi_{\mu}}, e^{i\phi_{\tau}})m_{\nu}diag(e^{i\phi_{e}}, e^{i\phi_{\mu}}, e^{i\phi_{\tau}}).}  \qquad 
\end{equation}
\end{small}
This entitles the freedom  to construct the quartet combination of the elements of $\rm{m_{\nu}}$ \cite{33}, which is rephasing invariant. In general, it is given by   
\begin{equation}\label{eq15}
\rm{Q\equiv m_{\alpha\beta}^{2}m_{\alpha\alpha}^{*}m_{\beta\beta}^{*}},
\end{equation}
  where, $\rm{\alpha,\beta}$ run over $\rm{e,\mu,\tau}$ ($\rm{\alpha\neq \beta}$). 
 The CP violation depends on the imaginary part of Q, and its expression can be given in terms of the CP violating phases, $ \rho, \sigma, \delta$, using  the expressions of $\rm{m_{ij}}$ (i, j=e, $\mu, \tau$). Hence, ImQ=0, implies CP invariance condition in leptonic sector.   This condition  does not imply that $\delta$, $\rho , \sigma$ are  necessarily zero, though it  could serve a purpose  to isolate them. 
 \\
 The analysis also stresses that quartets corresponding to all the seven viable ansatz of $m_{\nu}$ are nothing but a different  combination of Eq.(\ref{eq15}). 
This makes me to categorize the seven viable ansatze  into three types [ Table \ref{tab1}]. Apparently,   four ans\"atze (\rm{\textbf{{A1, A2, B3, B4}}})  belong to the quartet of the form, given in Eq.(\ref{eq15}), are grouped into  type I, while remaining ans\"atze, namely,  \rm{\textbf{{B1, B2}}}, and \textbf{C} are derived  form of the Q and are categorized as type II and type III, respectively.\\

\begin{table}[htp]
\begin{center}
\begin{tabular}{|c|c|c|}
\hline & \textbf{A1} & \textbf{A2}\\   
 \hline Type I &$\left(
\begin{array}{ccc}
0 & 0 & m_{e\tau} \\
 & m_{\mu\mu} & m_{\mu\tau}\\
&  & m_{\tau\tau} \\
\end{array}
\right)$ &$\left(
\begin{array}{ccc}
0 & m_{e\mu} &0  \\
 & m_{\mu\mu} & m_{\mu\tau}\\
&  & m_{\tau\tau} \\
\end{array}
\right)$ \\
\hline & \textbf{B3} & \textbf{B4} \\
\hline  &$\left(
\begin{array}{ccc}
    m_{ee}& 0 & m_{e\mu}\\
   & 0 & m_{\mu\tau}\\
 &  &m_{\tau\tau} \\ 
  \end{array}
\right)$
 &$\left(
\begin{array}{ccc}
     m_{ee}& m_{e\mu} & 0\\
   & m_{\mu\mu} & m_{\mu\tau}\\
 &  &0 \\
\end{array}
\right)$\\ \hline
 &\textbf{B1} & \textbf{B2}  \\
\hline Type II &$\left(
\begin{array}{ccc}
    m_{ee}& m_{e\mu} & 0 \\
   & 0 & m_{\mu\tau}\\
 &  & m_{\tau\tau}\\
\end{array}
\right)$& $\left(
\begin{array}{ccc}
    m_{ee}& 0 & m_{e\mu}\\
   & m_{\mu\mu} & m_{\mu\tau}\\
 &  &0 \\
\end{array}
\right)$\\ 
\hline
&$\textbf{C}$ &   \\
\hline Type III &$\left(
\begin{array}{ccc}
     m_{ee}& m_{e\mu} & m_{e\tau}\\
  & 0 & m_{\mu\tau}\\
  &  & 0 \\
\end{array}
\right)$&\\
\hline
\end{tabular}
\caption{\label{tab1}The seven viable ans\"{a}tze  of Majorana neutrino mass matrices having two texture zeros . }
\end{center}
\end{table}
In this section, I shall be stressing on how to isolate the CP violating phases $\rho$, $\sigma, \delta$ using imaginary part of the quartet following certain assumptions.  Earlier, R. Samanta et al.\cite{35}, have carried out an analysis to compute the Majorana phases  in model independent way for generic neutrino mass matrix using the three rephasing invariant quantities,  proposed by Sarkar and Singh \cite{20}. The analytical discussion is as follows\\

 \subsection{Type I}
 
 The invariant $\rm{I_{1}}$ can be derived using Eq.(\ref{eq4}), for ans\"atz $\textbf{A1}$ 
\begin{equation}\label{eq16}
\rm{I_{1}=-6i\Delta_{e\mu}\Delta_{\mu\tau}\Delta_{\tau e}|m_{e\tau}|^{2}ImQ_{A1}},
\end{equation}
where, $ \rm{Q_{A1}\equiv m_{\mu\mu}m_{\tau\tau}(m^{*}_{\mu\tau})^{2}}$, is the quartet for  ans\"atz $\textbf{A1}$ . The imaginary part of $\rm{Q_{A1}}$ is responsible for the leptonic CP violation.  \\  
The quartet can be expanded as a function of seven experimentally feasible neutrino parameters. However, it is not possible to separate the three CP violating phases without certain assumptions. To this end, I shall first derive the expression for  the imaginary part of $\rm{Q_{A1}}$. 
For that purpose, it is useful to divide $\rm{Q_{A1}}$ into two parts, in the leading order term of $s_{13}$ 
\begin{eqnarray}\label{eq17}
&&\rm{m_{\mu\mu}m_{\tau\tau}=a_{1}e^{4i(\rho-\delta)}+a_{2}e^{4i(\sigma-\delta)}+a_{3}e^{i(4\rho-3\delta)}
+a_{4}e^{i(4\sigma-3\delta)}+a_{5}e^{2i(\rho+\sigma-2\delta)}}
\nonumber  \\ 
 &&~~~~~~~~~~~~
\rm{+a_{6}e^{i(2\rho+2\sigma-3\delta)}+a_{7}e^{i(2\rho-\delta)}+a_{8}e^{i(2\sigma-\delta)}+a_{9}e^{2i(\rho-\delta)}}
\nonumber  \\ 
 &&~~~~~~~~~~~~
\rm{+a_{10}e^{2i(\sigma-\delta)}+a_{11}},
\end{eqnarray}
and,
\begin{eqnarray}\label{eq18}
&&\rm{(m_{\mu\tau}^{*})^{2}=a_{1}e^{-4i(\rho-\delta)}+a_{2}e^{-4i(\sigma-\delta)}+a_{3}e^{-i(4\rho-3\delta)}
+a_{4}e^{-i(4\sigma-3\delta)}+a_{5}e^{-2i(\rho+\sigma-2\delta)}}
\nonumber  \\ 
 &&~~~~~~~~~~~~
\rm{+a_{6}e^{-i(2\rho+2\sigma-3\delta)}+a_{7}e^{-i(2\rho-\delta)}+a_{8}e^{-i(2\sigma-\delta)}+a_{9}'e^{-2i(\rho-\delta)}}
\nonumber  \\ 
 &&~~~~~~~~~~~~
\rm{+a_{10}'e^{-2i(\sigma-\delta)}+a_{11}},
\end{eqnarray}
where, 
\begin{eqnarray*}
\rm{a_{1}=m_{1}^{2}c_{23}^{2}s_{23}^{2}s_{12}^{4}},\\
\rm{a_{2}=m_{2}^{2}c_{23}^{2}s_{23}^{2}c_{12}^{4}},\\
\rm{a_{3}=-2m_{1}^{2}c_{12}s_{12}^{3}c_{23}s_{23}s_{13}(c_{23}^{2}-s_{23}^{2})},\\
\rm{a_{4}=2m_{2}^{2}s_{12}c_{12}^{3}c_{23}s_{23}s_{13}(c_{23}^{2}-s_{23}^{2})},\\
\rm{a_{5}=2m_{1}m_{2}c_{23}^{2}s_{23}^{2}c_{12}^{2}s_{12}^{2}},\\
\rm{a_{6}=-2m_{1}m_{2}c_{12}s_{12}c_{23}s_{23}s_{13}(c_{12}^{2}-s_{12}^{2})(c_{23}^{2}-s_{23}^{2})},\\
\rm{a_{7}=2m_{1}m_{3}c_{12}s_{12}c_{23}s_{23}c_{12}s_{12}s_{13}c_{13}^{2}(c_{23}^{2}-s_{23}^{2})},\\
\rm{a_{8}=-2m_{2}m_{3}c_{12}s_{12}c_{23}s_{23}c_{12}s_{12}s_{13}c_{13}^{2}(c_{23}^{2}-s_{23}^{2})},\\
\rm{a_{9}=m_{1}m_{3}s_{12}^{2}c_{13}^{2}(c_{23}^{2}+s_{23}^{2})},\\
\rm{a_{10}=m_{2}m_{3}c_{12}^{2}c_{13}^{2}(c_{23}^{2}+s_{23}^{2})},\\
\rm{a_{9}'=-2m_{1}m_{3}c_{23}^{2}s_{23}^{2}s_{12}^{2}c_{13}^{2}},\\
\rm{a_{10}'=-2m_{2}m_{3}c_{23}^{2}s_{23}^{2}c_{12}^{2}c_{13}^{2}},\\
\rm{a_{11}=m_{3}^{2}s_{23}^{2}c_{23}^{2}c_{13}^{4}}.
\end{eqnarray*}
 are coefficients, depending on neutrino masses, $\rm{m_{1}, m_{2}, m_{3}}$, and neutrino mixing angles, $\rm{\theta_{12}, \theta_{23}, \theta_{13}}$.\\ 
Using Eqs.(\ref{eq17}) and (\ref{eq18}),  I get
\begin{small}
   \begin{eqnarray}\label{eq19}
&&\rm{ImQ_{A1}=\alpha_{1}sin(2\rho-\delta)+\alpha_{2}sin(2\sigma-\delta)+\alpha_{3}sin2(\rho-\delta)+\alpha_{4}sin2(\sigma-\delta)+\alpha_{5}sin2(\rho-\sigma)}
\nonumber  \\ 
 &&~~~~~~~~~~~~
 \rm{+\alpha_{6}sin\delta+\alpha_{7}sin2(2\rho-\sigma-\delta)+\alpha_{8}sin2(2\sigma-\rho-\delta)+\alpha_{9}sin(4\sigma-2\rho-\delta)}
 \nonumber  \\ 
 &&~~~~~~~~~~~~
 \rm{+\alpha_{10}sin(4\rho-2\sigma-\delta)+
\alpha_{11}sin(2\rho-2\sigma+\delta)
+\alpha_{12}sin(2\sigma-2\rho+\delta),}
\end{eqnarray}
 \end{small}
 where, the higher order terms, O($s_{13}^{2}$)  and more, are neglected. 
 The coefficients of sine terms are independent of $\rho,\sigma,\delta$, and are given as\
\begin{eqnarray*}
\rm{\alpha_{1}=a_{3}a_{9}'-a_{3}a_{9}+a_{6}a_{10}'-a_{6}a_{10},}\\
\rm{\alpha_{2}=a_{4}a_{10}'-a_{4}a_{10}+a_{6}a_{9}'-a_{6}a_{9},}\\
\rm{\alpha_{3}=a_{5}a_{10}'-a_{5}a_{10}+a_{1}a'_{9}-a_{1}a_{9},}\\
\rm{\alpha_{4}=a_{5}a_{9}'-a_{5}a_{9}+a_{2}a'_{10}-a_{2}a_{10},}\\
\rm{\alpha_{5}=a_{9}a_{10}'-a_{9}'a_{10},}\\
\rm{\alpha_{6}=a_{8}a_{10}'-a_{8}a_{10}+a'_{9}a_{7}-a_{9}a_{7},}\\
\rm{\alpha_{7}=a_{1}a_{10}'-a_{1}a_{10},}\\
\rm{\alpha_{8}=a_{2}a_{9}'-a_{2}a_{9},}\\
\rm{\alpha_{9}=a_{4}a_{9}'-a_{4}a_{9},}\\
\rm{\alpha_{10}=a_{3}a_{10}'-a_{3}a_{10},}\\
\rm{\alpha_{11}=a_{7}a_{10}'-a_{7}a_{10},}\\
\rm{\alpha_{12}=a_{8}a_{9}'-a_{8}a_{9}.}\\
\end{eqnarray*}
The remaining combinations of the sine terms are simply the complex conjugate of each other, and thus cancels out each other. The expression of Im$Q_{A1}$ shows that it is not possible to isolate the CP violating phases $\rho,\sigma,\delta$. In order to facilitate the computation of $\delta$, it is instructive to compute  $\rm{J_{CP}}$, which is sensitive to $\delta$ only.\\
With the help of Eqs.( \ref{eq9},\ref{eq16}), $\rm{J_{CP}}$ is expressed as
\begin{equation}\label{eq20}
\rm{J_{CP}=\frac{|m_{e\tau}|^{2}ImQ_{A_{1}}}{ \Delta_{sol} \Delta_{atm}^{2}}}.
\end{equation}
Since  $\delta$ is a only source of CP violation here, therefore, $\rho,\sigma=0$. Hence one can arrive at
\begin{small}
 \begin{eqnarray}\label{eq21}
 &&\rm{cos\delta=\frac{1}{4xv(y-z)}\big[-(xu(y-z)+v(x^{2}+(y-z)^{2}))}
 \nonumber  \\ 
 &&~~~~~
 \rm{\pm \sqrt{(xu(y-z)+v(x^{2}+(y-z)^{2}))^{2}-4xv(y-z)((x^{2}+(y-z)^{2})u+w)}\big]},
 \end{eqnarray}
 \end{small}
 following the Eqs.(\ref{eq19}) and (\ref{eq20}).  The symbols,  $\rm{x, y, z, u, v, w}$ are expressed below,
 \begin{eqnarray*}
 \rm{x=-m_{1}c_{13}s_{13}c_{12}^{2}c_{23}-m_{2}c_{13}s_{13}s_{12}^{2}c_{23}+m_{3}c_{13}s_{13}c_{23},} \\
  \rm{y=m_{1}c_{13}c_{12}s_{12}s_{23},} \\
  \rm{z=m_{2}c_{13}c_{12}s_{12}s_{23},}\\
  \rm{u= 2m_{3}(c_{23}^{2}-s_{23}^{2})c_{23}s_{23}c_{12}s_{12}s_{13}c_{13}^{2}[(m_{1}^{2}-m_{2}^{2}+2m_{1}m_{2}(c_{12}^{2}-s_{12}^{2}))}\\
\rm{(m_{1}s^{2}_{12}+m_{2}c^{2}_{12}) -(m_{1}s^{2}_{12}-m_{2}c^{2}_{12})(m_{1}+m_{2})-m_{1}m_{2}m_{3}s^{2}_{12}c_{13}^{2}],}\\
\rm{v= m_{3}c_{23}^{2}s_{23}^{2}c_{13}^{2}(m_{1}s^{2}_{12}+m_{2}c^{2}_{12})^{3},}\\
\rm{w=\frac{1}{8}s_{2(12)}s_{2(23)}s_{2(13)}c_{13}\Delta _{sol} \Delta _{atm}^{2}}.
\end{eqnarray*}
As a special case if CP invariance is considered, implying that $\rm{ImQ_{A_{1}}=0}$,\\ Eq. (\ref{eq20}) appears as 
\begin{equation}\label{eq22}
\rm{|m_{e\tau}|^{2}[(\alpha_{1}+\alpha_{2}+\alpha_{6}+\alpha_{9}+\alpha_{10}-\alpha_{11}-\alpha_{12})+2(\alpha_{3}+\alpha_{4}+\alpha_{7}\alpha_{8})cos\delta]sin\delta=0},
\end{equation}
where, $\rm{|m_{e\tau}|^{2}\neq 0}$ for ansatz \textbf{A1}, and $\rm{\Delta_{sol}}\neq 0$ and $ \rm{\Delta_{atm}^{2}}\neq 0$ are ruled out from the neutrino oscillation experiment.
 Therefore, I am left with   two possible solutions,
 \begin{equation}\label{eq23}
\rm{sin\delta=0,\qquad
 cos\delta=-\frac{(\alpha_{1}+\alpha_{2}+\alpha_{6}+\alpha_{9}+\alpha_{10}-\alpha_{11}-\alpha_{12})}{2(\alpha_{3}+\alpha_{4}+\alpha_{7}+\alpha_{8})}}.
\end{equation}
The cosine term can be further simplified in terms of the neutrino mass matrix parameters, $\rm{cos\delta=-\frac{u}{2v}}$.\\
From the above  expression, it is now feasible to compute the Dirac CP phase $\delta$ from WB invariants. In order to compute the  Majorana type phases, it is instructive to  find the  invariants  $\rm{I_{2}}$ and $\rm{I_{3}}$ since  they are  sensitive to  Majorana type CP phases besides Dirac type phase. 

%Using neutrino oscillation data at 3$\sigma$ confidence level (CL) for input parameters, $\phi=-26^{0}-26^{0}$, implying both CP violation and conservation.\\

%In Fig(), we have presented the correlation plot between  sin$\delta$ and sin$\phi$ and found that $\delta$ is restricted between -90-90.
Using Eq.(\ref{eq10}) and Eq.(\ref{eq12}),  $\rm{I_{2}}$ and $\rm{I_{3}}$ can be computed as
\begin{equation}\label{eq24}
\rm{I_{2}=\Delta_{\mu\tau}^{2}ImQ_{A1}},
\end{equation}
 and,
\begin{equation}\label{eq25}
\rm{I_{3}=6i\Delta_{e\mu}\Delta_{\mu\tau}\Delta_{\tau e} m_{\mu}^{2}m_{\tau}^{4} |m_{e\tau}|^{2}ImQ_{A1}}.
\end{equation}
 The computation of Majorana type phases is possible only if CP invariance condition is retained.  Taking it into account, while, using $\rm{sin\delta}=0$ from Eq.(\ref{eq23}), and $\rho=0$, in Eq.(\ref{eq24}), it is found 
\begin{equation}\label{eq26}
\rm{\Delta_{\mu\tau}^{2}[(\alpha_{2}+\alpha_{4}-\alpha_{5}-\alpha_{10}-\alpha_{11}+\alpha_{12})+2(\alpha_{8}+\alpha_{9})cos2\sigma]sin2\sigma=0}
\end{equation} 
where, $\Delta_{\mu\tau}^{2}\neq 0$. Therefore, there are two possible solutions either $\rm{sin2\sigma=0}$ or $\rm{cos2\sigma= -\frac{(\alpha_{2}+\alpha_{4}-\alpha_{5}-\alpha_{10}-\alpha_{11}+\alpha_{12})} {2(\alpha_{8}+\alpha_{9})
}}$.\\
On further simplification of coefficients $\alpha$'s in the cosine term, I  get
\begin{small}
\begin{eqnarray}\label{eq27}
&&\rm{cos2\sigma= \frac{1}{-m_{2}^{2}c_{12}^{3}s^{2}_{12}(c_{12}-2s_{12}s_{13}(s_{23}^{2}-c_{23}^{2}))}}
\nonumber  \\ 
 &&~~~~~
\times\rm{(-2m_{2}m_{3}c_{12}s_{12}s_{13}c_{13}^{2}(c_{23}^{2}-s_{23}^{2})}
\rm{(m_{1}^{2}s_{12}^{2}+m_{2}^{2}c_{12}^{2}+m_{1}^{3}(c_{12}^{2}-s_{12}^{2})
+m_{1}^{2}m_{2}c_{13}^{2})}
\nonumber  \\ 
 &&~~~~~
\rm{+m_{2}^{3}m_{3}c_{23}s_{23}c_{13}^{2}+2m_{1}m_{3}^{2}c_{12}s_{12}s_{13}(c_{23}^{2}-s_{23}^{2})-2m_{1}^{2}m_{2}m_{3}c_{23}s_{23}c_{12}^{2}s_{12}^{4})}.
 \end{eqnarray}
 \end{small}
Similarly, one can compute the expression for $\rho$, by assuming $\sigma=0$ in Eq.(\ref{eq24}),
\begin{equation}\label{eq28}
\rm{\Delta_{\mu\tau}^{2}[(\alpha_{1}+\alpha_{3}+\alpha_{5}-\alpha_{8}-\alpha_{9}+\alpha_{10}-\alpha_{12})+(\alpha_{7}+\alpha_{10})cos2\rho]sin2\rho=0},
\end{equation} 
which, again, gives rise to two possible solutions, either $\rm{sin2\rho=0}$ or \\$\rm{cos2\rho= -\frac{(\alpha_{1}+\alpha_{3}+\alpha_{5}-\alpha_{8}-\alpha_{9}+\alpha_{10}-\alpha_{12})} {2(\alpha_{7}+\alpha_{10})}}$.\\
Following the above procedure,  it is trivial to find the similar expressions for Majorana type phases  using $\rm{I_{3}=0}$ Eq.(\ref{eq25}). 
The results for ansatz \textbf{A2} can simply be obtained by replacing, $\rm{\theta_{23}^{0}}$ with $\rm{90^{0}-\theta_{23}^{0}}$, and \rm{$\delta^{0}$ with $180- \delta^{0}$}, in  ansatz \textbf{A1}.\\
 Similar to ansatze \textbf{A1, A2}, ansatze \textbf{B3, B4} also belong  to the family of quartet,  defined in Eq.(\ref{eq15}). This can be shown by deriving the WB invariant $\rm{I_{1}}$ using  Eq.(\ref{eq4}) 
\begin{equation}\label{eq29}
\rm{I_{1}=-6i\Delta_{e\mu}\Delta_{\mu\tau}\Delta_{\tau e}|m_{\mu\tau}|^{2} ImQ_{B_{3}}}
\end{equation}
where, $\rm{Q_{B3}\equiv (m_{e\tau}^{2})^{*}m_{ee}m_{\tau\tau}}$, is the quartet for ansatz \textbf{B3}. The imaginary part of $\rm{Q_{B3}}$ is responsible for the leptonic CP violation.  
  
The explicit expression of  imaginary part of  $\rm{Q_{B3}}$ can be computed in terms of $\delta, \rho,
\sigma$. To this end, first of all, I express the complex, and complex conjugate part of  $\rm{Q_{B3}}$
, in the leading order term of $s_{13}$
\begin{small}
   \begin{eqnarray*}
&&\rm{m_{ee}m_{\tau\tau}=b_{1}e^{2i(2\rho-\delta)}+b_{2}e^{2i(2\sigma-\delta)}+b_{3}e^{i(4\rho-\delta)}+b_{4}e^{i(4\sigma-\delta)}+b_{5}e^{i(2\rho+ 2\sigma-\delta)}}
\nonumber  \\ 
 &&~~~~~~~~~~~~~~~~
\rm{+b_{6}e^{2i(\rho +\sigma -\delta)}+b_{7}e^{2i(\sigma)}+b_{8}e^{2i(\rho)}}
\end{eqnarray*}
\end{small}
and,
\begin{small}
   \begin{eqnarray*}
&&\rm{(m_{e\tau}^{*})^{2}=b_{1}e^{-2i(2\rho-\delta)}+b_{2}e^{-2i(2\sigma-\delta)}+b_{3}e^{-i(4\rho-\delta)}+b_{4}e^{-i(4\sigma-\delta)}+b_{5}e^{-i(2\rho+ 2\sigma-\delta)}}
\nonumber  \\ 
 &&~~~~~~~~~~~~~~~~
\rm{+b_{6}'e^{-2i(\rho +\sigma -\delta)}+b_{7}'e^{-i(2\rho-\delta)}+b_{8}'e^{-i(2\sigma-\delta)},}
\end{eqnarray*}
\end{small}
 where, the coefficients  are the  function of neutrino masses and mixing angles only, and are given as, $ \rm{b_{1}=m_{1}^{2}c_{12}^{2}s_{12}^{2}s_{23}^{2}c_{13}^{2}}$, \quad $ \rm{ b_{2}=m_{2}^{2}c_{12}^{2}s_{12}^{2}s_{23}^{2}c_{13}^{2}}$,\\
 \quad $\rm{b_{3}=-2m_{1}^{2}c_{12}^{3}s_{12}c_{23}s_{23}c_{13}^{2}s_{13}}$, \quad $\rm{b_{4}=2m_{2}^{2}s_{12}^{3}c_{12}c_{23}s_{23}c_{13}^{2}s_{13}}$,\\
 \quad $\rm{b_{5}=2m_{1}m_{2}c_{12}s_{12}c_{23}s_{23}c_{13}^{2}s_{13}(c_{12}^{2}-s_{12}^{2})}$, \quad
$\rm{b_{6}=m_{1}m_{2}(c_{12}^{4}+s_{12}^{4})s_{23}^{2}c_{13}^{2}}$,\\
\quad $\rm{b_{7}=m_{2}m_{3}s_{12}^{2}c_{23}^{2}c_{13}^{4}}$, \quad
$\rm{b_{8}=m_{1}m_{3}c_{12}^{2}c_{23}^{2}c_{13}^{4}}$, \quad\\
$\rm{b_{6}'=m_{1}m_{2}c_{13}^{2}s_{13}c_{12}^{3}s_{12}c_{23}s_{23}}$,\quad
$\rm{b_{7}'=-2m_{2}m_{3}c_{13}^{2}s_{13}c_{12}s_{12}s_{23}c_{23}}$,\quad\\
$\rm{b_{8}'=2m_{1}m_{3}c_{13}^{2}s_{13}c_{12}s_{12}s_{23}c_{23}}.
$

After carrying out the tedious calculations, I finally arrive at
\begin{small}
   \begin{eqnarray}\label{eq30}
&& \rm{ImQ_{B3}=\xi_{1}sin2(\rho-\sigma)+\xi_{2}sin(2\rho-\delta)+ \xi_{3}sin(2\sigma-\delta)+ \xi_{4}sin2(2\rho-\sigma-\delta)}
\nonumber  \\ 
 &&~~~~~~~~
 \rm{+\xi_{5}sin2(2\sigma-\rho-\delta) +\xi_{6}sin2(\rho-\delta)+ \xi_{7}sin2(\sigma-\delta)+ \xi_{8}sin2\rho}
 \nonumber  \\ 
 &&~~~~~~~~
\rm{ + \xi_{9}sin2\sigma
 +\xi_{10}sin(4\rho-2\sigma-\delta)+ \xi_{11}sin(4\sigma-2\rho-\delta) +\xi_{12}sin(2\rho-2\sigma+ \delta)}  
 \nonumber  \\ 
 &&~~~~~~~~
\rm{+ \xi_{13}sin(2\sigma-2\rho +\delta)  +\xi_{14}sin2(2\rho-\delta) 
 +\xi_{15}sin2(2\sigma-\delta) + \xi_{16}sin\delta.}
\end{eqnarray}
\end{small}
where, $\xi_{i} $(where,$i=1 - 16$), represents the term containing neutrino masses and mixing angles , and are  independent of $\rho,\sigma,\delta$. \\
\begin{eqnarray*}
\rm{\xi_{1}=b_{1}b_{6}'-b_{1}b_{6}- b_{2}b_{6}'+b_{2}b_{6}},\\
\rm{\xi_{2}=b_{1}b_{7}'-b_{8}b_{3}- b_{7}b_{5}-b_{7}b_{6}'},\\
\rm{\xi_{3}=b_{2}b_{8}'-b_{7}b_{4}- b_{8}b_{5}+b_{6}b_{7}'},\\
\rm{\xi_{4}=b_{7}b_{1},}\\
\rm{\xi_{5}=-b_{8}b_{2},}\\
\rm{\xi_{6}=-b_{8}b_{1},}\\
\rm{\xi_{7}=-b_{7}b_{2},}\\
\rm{\xi_{8}=b_{3}b'_{7}+b_{5}b_{8}',}\\
\rm{\xi_{9}=b_{4}b'_{8}+b_{5}b_{7}',}\\
\rm{\xi_{10}=b_{1}b'_{8}-b_{7}b_{3},}\\
\rm{\xi_{11}=b_{2}b'_{7}-b_{8}b_{4},}\\
\rm{\xi_{12}=b_{3}b_{6}'-b_{3}b_{6}- b_{7}b_{7}',}\\
\rm{\xi_{13}=b_{4}b_{6}'-b_{4}b_{6}- b_{8}b_{8}',}\\
\rm{\xi_{14}=b_{3}b_{8}',}\\
\rm{\xi_{15}=b_{4}b'_{7},}\\
\rm{\xi_{16}=b_{5}b_{6}'-b_{5}b_{6}+b_{8}b_{7}'+b_{7}b_{8}'}.\\
\end{eqnarray*}
Eq.(\ref{eq30}), shows the  dependence on the possible combinations of CP phases. Looking at the above relation, it is difficult to isolate the CP phases unless some assumptions are made.
Following the same procedure as in case of ansatz \textbf{A1},  $\delta$ can be computed. \\
The Jarlskog rephasing invariant, $\rm{J_{CP}}$ can be computed as
\begin{equation}\label{eq31}
\rm{J_{CP}=\bigg[\frac{|m_{\mu\tau}|^{2} ImQ_{B_{3}}}{\Delta_{sol} \Delta_{atm}^{2}}\bigg]}.
\end{equation}
As already pointed out,  $\rm{I_{1}}$ is sensitive to  $\delta$ only. Therefore  Majorana type phases are automatically vanishing, i.e., $\rho,\sigma=0$. \\
In the approximation, $m_{1}\simeq m_{2}$, Eq.(\ref{eq31}) reduces to 
\begin{eqnarray}\label{eq32}
\rm{[c+dcos2\delta]
[b+2acos\delta]sin\delta= w sin\delta},
\end{eqnarray}
where, coefficients \rm{a, b, c, d},  are independent of $\delta$, and are given as\\
\rm{$a= -(\xi_{4}+\xi_{5})$,} \rm{$b=(-\xi_{2}-\xi_{3}-\xi_{6}-\xi_{10}-\xi_{11}-\xi_{12}-\xi_{13}+\xi_{16})$},\\
\rm{$c=1+m_{3}^{2}s^{2}_{23}c^{2}_{23}c^{4}_{13}$,} \rm{ $d=-2m_{1}m_{3}c^{2}_{23}s^{2}_{23}c^{2}_{13}$}.\\
On simplification, Eq.(\ref{eq32}) takes the  cubic form ,
 \begin{eqnarray}\label{eq33}
\rm{Acos^{3}\delta+Bcos^{2}\delta +Ccos\delta+ D =0,}
\end{eqnarray}
where, A=4ad, B=2bd, C=2a(c-d), D=b(c-d)-w.\\
Applying Cardano's formulation, it is possible to solve the above cubic equation. With the help of $\rm{cos\delta=\lambda-\frac{B}{3}}$, Eq.(\ref{eq33}) is reduced to 
\begin{equation}\label{eq34}
\rm{\lambda^{3}+P\lambda+Q=0},
\end{equation}
where, $\rm{P=\big(C-\frac{B^{2}}{3}\big), Q=\big(\frac{2B^{3}}{27}-\frac{BC}{3}+D\big)}$.\

Inserting $\rm{\lambda=\kappa-\frac{P}{3\kappa}}$ in  Eq.(\ref{eq34}) reduces it to  quadratic form in terms of the variable $\kappa^{3}$, and its solution is given as $\rm{\kappa=(-\frac{Q}{2}+r)^{1/3}}$, where, $\rm{r=\pm\sqrt{(\frac{Q^{2}}{4}+\frac{P^{3}}{27})}}$.\\ 
Finally solutions for  $\delta$, is provided below
\begin{equation}\label{eq35}
\rm{cos\delta= \big(\kappa-\frac{P}{3\kappa}\big)-\frac{B}{3}}.
\end{equation}
Therefore, adopting the cardano's formulation, it is possible to compute $\delta$ for ansatz \textbf{B3}.\\
Alternatively, following the CP invariance condition, ( i.e., $\rm{ImQ_{B_{3}}=0}$),  Eq.(\ref{eq29}) reduces to, 
\begin{eqnarray}\label{eq36}
&&\rm{\rm{\Delta_{e\mu}\Delta_{\mu\tau}\Delta_{\tau e}|m_{\mu\tau}|^{2}[(-\xi_{2}-\xi_{3}-\xi_{6}-\xi_{10}-\xi_{11}-\xi_{12}-\xi_{13}+\xi_{16})}}
\nonumber  \\ 
 &&~~~~~~~~
\rm{\rm{-2(\xi_{4}+\xi_{5})cos\delta]sin\delta=0}},
\end{eqnarray}
where, $\Delta_{ij}\neq 0$, and $m_{\mu\tau}\neq 0$ follows from ansatz \textbf{B3}. Hence Eq.(\ref{eq36}), implies  two possible solutions, 
\begin{equation}\label{eq37}
\rm{ sin\delta=0, \quad cos\delta= \frac{(-\xi_{2}-\xi_{3}-\xi_{6}-\xi_{10}-\xi_{11}-\xi_{12}-\xi_{13}+\xi_{16})}{2(\xi_{4}+\xi_{5})}},
\end{equation}
where, symbols have their usual meaning.
The invariants $\rm{I_{2}}$ and $\rm{I_{3}}$ can be computed as
\begin{equation}\label{eq38}
\rm{I_{2}=\Delta_{\mu\tau}^{2}ImQ_{B_{3}}},
\end{equation}
 and,
\begin{equation}\label{eq39}
\rm{I_{3}=6i\Delta_{e\mu}\Delta_{\mu\tau}\Delta_{\tau e} m_{\mu}^{2}m_{\tau}^{4}ImQ_{B_{3}},},
\end{equation}
respectively.
Following  sin$\delta=0$ from Eq.(\ref{eq37}), and assuming $\rho=0$,  CP invariance condition in Eq.(\ref{eq38}) lead to
\begin{eqnarray}\label{eq40}
&&\rm{\Delta_{\mu\tau}^{2}[(\xi_{1}+\xi_{2}-\xi_{5}+ \xi_{6}+\xi_{8}-\xi_{11}+\xi_{12}-\xi_{13})}
\nonumber  \\ 
 &&~~~~~~~~
\rm{+2(\xi_{4}+\xi_{10}+\xi_{14})cos2\sigma]sin2\sigma=0.}
\end{eqnarray}
The above equation yields two solutions,
\begin{equation}\label{eq41}
\rm{sin2\sigma=0, \quad cos2\sigma= -\frac{(\xi_{1}+\xi_{2}-\xi_{5}+ \xi_{6}+\xi_{8}-\xi_{11}+\xi_{12}-\xi_{13})}{2(\xi_{4}+\xi_{10}+\xi_{14})}}.
\end{equation} 
 
Similarly, assuming $\sigma=0$, CP invariance condition in Eq.(\ref{eq38}) lead to
\begin{eqnarray}\label{eq42}
&& \rm{\Delta_{\mu\tau}^{2}[(-\xi_{1}-\xi_{4}+\xi_{7}+\xi_{9}-\xi_{10}-\xi_{12}+\xi_{13})}
\nonumber  \\ 
 &&~~~~~~~~
\rm{+2(\xi_{5}+\xi_{11}+\xi_{15})cos2\rho]sin2\rho=0.}
\end{eqnarray}
The equation again leads to two solutions,
 \begin{equation}\label{eq43}
\rm{ sin2\rho=0, \quad cos2\rho= \frac{(-\xi_{1}-\xi_{4}+\xi_{7}+\xi_{9}-\xi_{10}-\xi_{12}+\xi_{13})}{2(\xi_{5}+\xi_{11}+\xi_{15})}}.
 \end{equation}
Similar expressions for $\rho$ and $\sigma$ can also be obtained  by employing the CP invariance condition in $\rm{I_{3}}$. 
The phenomenological results for ans\"atz \textbf{B4} can be obtained by the exchange of $\mu-\tau$ symmetry.\\
\subsection{\rm{Type II}}
The WB invariant $\rm{I_{1}}$ can be  computed using Eq.(\ref{eq4}),
\begin{equation}\label{eq44}
\rm{I_{1}=-6i\Delta_{e\mu}\Delta_{\mu\tau}\Delta_{\tau e}ImQ_{B_{1}}},
\end{equation}
 where, $\rm{Q_{B1}\equiv ((m_{\mu \tau}^{*})^{2}m_{ee}^{*}m_{e\mu}^{2}m_{\tau\tau})}$, and the imaginary part of $\rm{Q_{B1}}$ is a measure of CP violation. \\
 The  quartet of ans\"atz \textbf{B1} can be derived from type I,  \\
 \begin{equation}\label{eq45}
 \rm{Q_{B1}^{*}=|m_{\mu\tau}^{2}|^{2}|m_{\mu\mu}|^{2}\frac{Q_{B4}}{Q_{A1}}}.
 \end{equation}
 Similarly, for ansatz \textbf{B2}, I have, $\rm{Q_{B2}^{*}=|m_{\mu\tau}^{2}|^{2}|m_{\tau\tau}|^{2}\frac{Q_{B3}}{Q_{A1}}}$.\\
 From these relations, one find that $\rm{argQ_{A1}-argQ_{B4}=argQ_{B1}}$. Similarly, one can get, $\rm{argQ_{A1}-argQ_{B3}=argQ_{B2}}$. Using these relations, I have
 \begin{equation*}
 \rm{arg\frac{Q_{B4}}{Q_{B3}}= arg \frac{Q_{B1}}{Q_{B2}}}.
 \end{equation*}
 The relations suggest that relative phase angles of ansatze \textbf{B} are not independent of each other.\\
  Using Eq.(\ref{eq45}), the imaginary part of $\rm{Q_{B1}}$ can be explicitly derived in terms of $\rho,\sigma,\delta$.\\
The $\rm{J_{CP}}$ can be computed through the relation
\begin{equation}\label{eq46}
\rm{J_{CP}=\bigg[\frac{ImQ_{B1}}{\Delta_{sol} \Delta_{atm}^{2}}\bigg]}.
\end{equation}
The WB invariants $\rm{I_{2}}$ is found to be exactly zero \ref{eq11}
\begin{equation}\label{eq47}
\rm{I_{2}=0}.
\end{equation}
 This is because in the expression of $\rm{I_{2}}$, it is found that there is a exact cancellation of phases associated with the elements, $\rm{m_{ee},m_{e\mu}, m_{\mu\tau},m_{\tau\tau}}$ since the conjugate pairs of the same is found.\\
 The expression of $\rm{I_{3}}$ can be obtained as \ref{eq13}
\begin{equation}\label{eq48}
\rm{I_{3}=6i\Delta_{e\mu}\Delta_{\mu\tau}\Delta_{\tau e}m_{e}^{2}m_{\mu}^{2}m_{\tau}^{2}ImQ_{B1}.}
\end{equation}
The results for $B_{2}$ can be obtained by simply using $\mu-\tau$ permutation symmetry. 
\subsection{\rm{ Type III}}

The WB invariant $I_{1}$ is given as
\begin{equation}\label{eq49}
\rm{I_{1}=-6i\Delta_{e\mu}\Delta_{\mu\tau}\Delta_{\tau e}(|m_{e\mu}|^{2}-|m_{e\tau}|^{2})ImQ_{C}},
\end{equation}
where, $\rm{Q_{C}\equiv m_{\mu\tau}m_{ee}m_{e\tau}^{*} m_{e\mu}^{*}}$ is a  quartet in ansatz \textbf{C}, and imaginary part of $\rm{Q_{C}}$ is  responsible for CP violation.\\
Similar to  type II, the quartets  pertaining to type III can  be derived using the combinations of \rm{Q},    
\begin{equation}\label{eq50}
\rm{Q_{C}=\frac{1}{|m_{\mu\mu}||m_{\tau\tau}|}\sqrt{Q_{B_{3}}Q_{B_{4}}Q_{A_{1}}^{*}}},
\end{equation}
where, $\rm{Q_{C}}$ denotes the quartet in ans\"atz \textbf{C}. \\
In the leading order approximation of $s_{13}$,
 $\rm{Q_{C}}$ can be divided into a complex part and its conjugate part 
\begin{small}
   \begin{eqnarray}\label{eq51}
&&\rm{m_{ee}m_{\tau\tau}=c_{1}e^{i(4\rho-\delta)}+c_{2}e^{i(4\sigma-\delta)}+c_{3}e^{2i(2\rho-\delta)}+c_{4}e^{2i(2\sigma-\delta)}+c_{5}e^{i(2\rho+ 2\sigma-\delta)}}
\nonumber  \\ 
 &&~~~~~~~~~~~~~~~~
\rm{+c_{6}e^{2i(\rho +\sigma -\delta)}+c_{7}e^{2i(\sigma)}+c_{8}e^{2i(\rho)}},
\end{eqnarray}
\end{small}
and,
\begin{small}
   \begin{eqnarray}\label{eq52}
&&\rm{(m_{e\tau}^{*})^{2}=c_{1}e^{-i(4\rho-\delta)}+c_{2}e^{-i(4\sigma-\delta)}+c_{3}e^{-2i(2\rho-\delta)}+c_{4}e^{-2i(2\sigma-\delta)}+c_{5}e^{-i(2\rho+ 2\sigma-\delta)}}
\nonumber  \\ 
 &&~~~~~~~~~~~~~~~~
\rm{+c_{6}'e^{-2i(\rho +\sigma -\delta)}+c_{7}'e^{-i(2\rho-\delta)}+c_{8}'e^{-i(2\sigma-\delta)}},
\end{eqnarray}
\end{small}
 where, the coefficients  are the  function of neutrino masses and mixing angles, and are given as
 \begin{eqnarray*}
\rm{c_{1}=m_{1}^{2}c_{12}^{3}s_{12}c_{13}^{2}s_{13} (c_{23}^{2}-s_{23}^{2})},\\
\rm{c_{2}=m_{2}^{2}s_{12}^{3}c_{12}c_{13}^{2}s_{13} (c_{23}^{2}-s_{23}^{2}),}\\
\rm{c_{3}=-m_{1}^{2}c_{12}^{2}s_{12}^{2}c_{23}s_{23}c_{13}^{2},}\\
\rm{c_{4}=-m_{2}^{2}c_{12}^{2}s_{12}^{2}c_{23}s_{23}c_{13}^{2},}\\
\rm{c_{5}=m_{1}m_{2}c_{13}^{2}s_{13}c_{12}s_{12}(s_{12}^{2}-c_{12}^{2})(c_{23}^{2}-s_{23}^{2}),}\\
\rm{c_{6}=-m_{1}m_{2}c_{13}^{2}s_{13}c_{23}s_{23}(c_{12}^{4}+s_{12}^{4}),}\\
\rm{c_{7}=m_{2}m_{3}s_{12}^{2}c_{23}s_{23}c_{13}^{4},}\\
\rm{c_{8}=m_{1}m_{3}c_{12}^{2}c_{23}s_{23}c_{13}^{4},}\\
\rm{c_{6}'=2c_{23}s_{23}s_{12}^{2}c_{12}^{2}c_{13}^{2},}\\
\rm{c_{7}'=m_{2}m_{3}c_{13}^{2}s_{13}c_{12}s_{12}(c_{23}^{2}-s_{23}^{2}),}\\
\rm{c_{8}'=-m_{1}m_{3}c_{13}^{2}s_{13}c_{12}s_{12}(c_{23}^{2}-s_{23}^{2})}.
\end{eqnarray*}
 After carrying out the tedious calculations, it is found that,
\begin{small}
   \begin{eqnarray}\label{eq53}
&& \rm{ImQ_{C}=\epsilon_{1}sin2(\rho-\sigma)+\epsilon_{2}sin(2\rho-\delta)+ \epsilon_{3}sin(2\sigma-\delta)+ \epsilon_{4}sin2(2\rho-\sigma-\delta)}
\nonumber  \\ 
 &&~~~~~~~~
 \rm{+\epsilon_{5}sin2(2\sigma-\rho-\delta) +\epsilon_{6}sin2(\rho-\delta)+ \epsilon_{7}sin2(\sigma-\delta)+ \epsilon_{8}sin2\rho}
 \nonumber  \\ 
 &&~~~~~~~~
 \rm{+ \epsilon_{9}sin2\sigma
 +\epsilon_{10}sin(4\rho-2\sigma-\delta)+ \epsilon_{11}sin(4\sigma-2\rho-\delta) +\epsilon_{12}sin(2\rho-2\sigma+ \delta)}  
 \nonumber  \\ 
 &&~~~~~~~~
\rm{+ \epsilon_{13}sin(2\sigma-2\rho +\delta)  +\epsilon_{14}sin2(2\rho-\delta) 
 +\epsilon_{15}sin2(2\sigma-\delta) + \epsilon_{16}sin\delta.}
\end{eqnarray}
\end{small}
where, \rm{$\epsilon_{i}$}( i=1 - 16), represents the term containing neutrino masses and mixing angles , and are  independent of $\rm{\rho},\sigma,\delta$. The remaining combinations are cancelled out due to the exact complex conjugate pair terms. In addition, it is explicit that  sine terms are similar to ans\"atz \textbf{B3}, however $\rm{\epsilon_{i}} $  are different, and are given as
 \begin{eqnarray*}
\rm{\epsilon_{1}=c_{3}c_{6}'-c_{3}c_{6}-c_{4}c_{6}'+c_{4}c_{6},}\\
\rm{\epsilon_{2}=c_{3}c_{8}'+c_{6}c_{7}'-c_{8}c_{1}-c_{7}c_{5},}\\
\rm{\epsilon_{3}=c_{4}c_{8}'+c_{6}c_{8}'-c_{7}c_{2}-c_{8}c_{5},}\\
\rm{\epsilon_{4}=c_{7}c_{5},}\\
\rm{\epsilon_{5}=-c_{8}c_{4},}\\
\rm{\epsilon_{6}=-c_{8}c_{3}-c_{7}c_{6}',}\\
\rm{\epsilon_{7}=-c_{7}c_{4}-c_{8}c_{6}',}\\
\rm{\epsilon_{8}=c_{7}c_{8}'+c_{5}c_{7}',}\\
\rm{\epsilon_{9}=c_{2}c'_{8}+c_{5}c'_{8},}\\
\rm{\epsilon_{10}=c_{3}c_{7}'-c_{7}c_{1},}\\
\rm{\epsilon_{11}=c_{4}c_{7}'-c_{8}c_{2},}\\
\rm{\epsilon_{12}=c_{1}c_{6}'-c_{1}c_{6}+c_{8}c_{7}',}\\
\rm{\epsilon_{13}=c_{2}c_{6}'-c_{6}c_{2}-c_{7}c'_{8},}\\
\rm{\epsilon_{14}=c_{1}c_{7}'},\\
\rm{\epsilon_{15}=c_{2}c_{7}',}\\
\rm{\epsilon_{16}=c_{5}c_{6}'-c_{5}c_{6}+c_{7}c_{7}'+c_{8}c_{8}'.}
\end{eqnarray*}
The Jarlskog rephasing invariant $\rm{J_{CP}}$ can be given   as
\begin{equation}\label{eq54}
\rm{J_{CP}=\bigg[\frac{(|m_{e\mu}|^{2}-|m_{e\tau}|^{2})ImQ_{C}}{\Delta_{sol} \Delta_{atm}^{2}}\bigg]}.
\end{equation}
The relation also implies that $\rm{I_{1}}$ is sensitive to $\rm{\delta}$ only.
On solving Eq.(\ref{eq54}), one finds
\begin{small}
\begin{eqnarray}\label{eq55}
&&\rm{s[t-2(\epsilon_{4}+\epsilon_{5})cos\delta]sin\delta =w sin\delta},
\end{eqnarray}
\end{small}
where, s and t are the coefficients depending on neutrino masses and mixing angles, 
$\rm{s=[c_{13}^{2}s_{13}^{2}(m_{1}c_{12}^{2}+m_{2}s_{12}^{2})+2c_{13}^{2}s_{13}^{2}(m_{1}c_{12}^{2}+m_{2}s_{12}^{2})(c_{23}^{2}-s_{23}^{2})
+m_{3}^{2}c_{13}^{2}s_{13}^{2}(s_{23}^{2}-c_{23}^{2})
]}$,
 $\rm{t=(-\epsilon_{2}-\epsilon_{3}-\epsilon_{6}-\epsilon_{10}-\epsilon_{11}-\epsilon_{12}-\epsilon_{13}+\epsilon_{16})}$.
 
 On simplification,  Eq.(\ref{eq54}) leads to
 \begin{equation}\label{eq56}
 \rm{cos\delta=\frac{st-w}{2s(\epsilon_{4}+\epsilon_{5})}}.
 \end{equation}
 Again as a special case if 
 CP invariance  is assumed, i.e., $\rm{I_{1}}=0$
  , one finds that either
 $\rm{m_{e\mu}=m_{e\tau}}$ , which means $\rm{\mu-\tau}$ symmetry since $\rm{m_{\mu\mu}=m_{\tau\tau}}$ holds for type III or otherwise one would have
\begin{equation}\label{eq57}
[(-\epsilon_{2}-\epsilon_{3}-\epsilon_{6}-\epsilon_{10}-\epsilon_{11}-\epsilon_{12}-\epsilon_{13}+\epsilon_{16})-2(\epsilon_{4}+\epsilon_{5})cos\delta]sin\delta=0
\end{equation}
where,  $\Delta_{ij}=0$  are avoided since they cannot be zero.\\

The above equation again leads to  the two possible solutions , i.e., sin$\delta=0$, 
or 
\begin{equation}\label{eq58}
 \rm{cos\delta= \frac{(-\epsilon_{2}-\epsilon_{3}-\epsilon_{6}-\epsilon_{10}-\epsilon_{11}-\epsilon_{12}-\epsilon_{13}+\epsilon_{16})}{2(\epsilon_{4}+\epsilon_{5})}}.
\end{equation}
Before computing Majorana type phases,  I, first of all, derive  the expression for invariants $\rm{I_{2}}$ and $\rm{I_{3}}$ using Eqs. (\ref{eq11}, \ref{eq13})
\begin{equation}\label{eq59}
\rm{I_{2}=2\Delta_{e\mu}\Delta_{\tau e}ImQ_{C}},
\end{equation}
and,
\begin{equation}\label{eq60}
\rm{I_{3}=6i\Delta_{e\mu}\Delta_{\mu\tau}\Delta_{\tau e} m_{e}^{4}(m_{\mu}^{2}|m_{e\mu}|^{2}-m_{\tau}^{2}|m_{e\tau}|^{2})ImQ_{C}}.
\end{equation}
 Assuming sin$\delta=0$ and CP invariance condition,  Eq.(\ref{eq59}) after putting $\rho=0$, \\
 lead to the following equation,
\begin{equation}\label{eq61}
\rm{2\Delta_{e\mu}\Delta_{\tau e}[(\epsilon_{1}+\epsilon_{2}-\epsilon_{5}+ \epsilon_{6}+\epsilon_{8}-\epsilon_{11}+\epsilon_{12}-\epsilon_{13})+2(\epsilon_{4}+\epsilon_{10}+\epsilon_{14})cos2\sigma]sin2\sigma=0}.
\end{equation}
 which implies,  
 \begin{equation}\label{eq62}
  \rm{sin2\sigma=0, \quad \quad cos2\sigma= -\frac{(\epsilon_{1}+\epsilon_{2}-\epsilon_{5}+ \epsilon_{6}+\epsilon_{8}-\epsilon_{11}+\epsilon_{12}-\epsilon_{13})}{2(\epsilon_{4}+\epsilon_{10}+\epsilon_{14})}}.
 \end{equation}
 
 Similarly, assuming $\rm{\sigma=0}$, Eq.(\ref{eq59}) lead to 
\begin{equation}\label{eq63}
\rm{[(-\epsilon_{1}-\epsilon_{4}+\epsilon_{7}+\epsilon_{9}-\epsilon_{10}-\epsilon_{12}+\epsilon_{13})+2(\epsilon_{5}+\epsilon_{11}+\epsilon_{15})cos2\rho]sin2\rho=0}.
\end{equation}
 The above equation leads to the two solutions 
 \begin{equation}\label{eq64}
 \rm{ sin2\rho=0, \quad cos2\rho= \frac{(-\epsilon_{1}-\epsilon_{4}+\epsilon_{7}+\epsilon_{9}-\epsilon_{10}-\epsilon_{12}+\epsilon_{13})}{2(\epsilon_{5}+\epsilon_{11}+\epsilon_{15})}}.
 \end{equation}
 
  In principle, similar results for Majorana type phases can be obtained by employing the CP invariance condition in Eq.(\ref{eq60}).
 To summarize the discussion, I have attempted to isolate the CP phases, and compute the expressions for CP  phases in terms of the observables of mass matrix for all the seven viable ans\"atz.\\

\begin{table}[ht]
\begin{small}
\begin{center}
\resizebox{15cm}{!}{

\begin{tabular}{|c|c|c|}
%  % after \\: \hline or \cline{col1-col2} \cline{col3-col4} ...
  \hline
Ans\"atz&Normal mass ordering
(NO) &Inverted mass ordering(IO) \\ 
\hline 
A1  & $I_{1}=
($-$1.104-91.106)\times 10^{-11}$ & $\times$  \\
 &$I_{2}=
($-$1.52-1.51)\times 10^{-6}$ &$\times$ \\
&$I_{3}=
($-$1.24-1.25)\times 10^{-12}$ & $\times$ \\
&$\Phi_{\rm{A1}}=$-$28.9^{0}-28.5^{0}$&$\times$ \\
\hline 
A2 & $I_{1}=
($-$1.104-1.105)\times 10^{-11}$ & $\times$  \\
 &$I_{2}=
($-$1.53-1.54)\times 10^{-6}$ &$\times$ \\
&$I_{3}=
($-$1.23-1.22)\times 10^{-12}$ & $\times$ \\
&$\Phi_{\rm{A2}}=$-$28.8^{0}-28.6^{0}$&$\times$\\
\hline
 B1 & $I_{1}=
($-$1.07\times 10^{-11}- $-$5.41\times 10^{-12})\oplus (5.57\times 10^{-12}-1.05\times 10^{-11})$& $I_{1}=
($-$1.05\times 10^{-11}- $-$5.58\times 10^{-12})\oplus (5.45\times 10^{-12}-1.05\times 10^{-11})$  \\
 &$I_{2}=0.0
$&$I_{2}=0.0$ \\
&$I_{3}=
($-$9.90\times 10^{-20}- $-$7.87 \times 10^{-20})\oplus (7.75\times 10^{-20}-9.88\times 10^{-20}) $ & $I_{3}=($-$9.70\times 10^{-20}- $-$7.67 \times 10^{-20})\oplus (7.61\times 10^{-20}-9.69\times 10^{-20})$
\\
&$\rm{\Phi_{B1}}=$-$(15^{0}-9.4^{0})\oplus (9.35^{0}-15^{0})$&$\rm{\Phi_{B1}}=$-$(172.15^{0}-162.05^{0})\oplus (162.4^{0}-171.98^{0})$\\
\hline
B2 & $I_{1}=
($-$1.05\times 10^{-11}- $-$5.49\times 10^{-12})\oplus (5.54\times 10^{-12}-1.07\times 10^{-11})$& $I_{1}=
($-$1.09\times 10^{-11}- $-$5.59\times 10^{-12})\oplus (5.45\times 10^{-12}-1.08\times 10^{-11})$  \\
 &$I_{2}=0.0
$&$I_{2}=0.0$ \\
&$I_{3}=
($-$9.76\times 10^{-20}- $-$7.89 \times 10^{-20})\oplus (7.78\times 10^{-21}-9.75\times 10^{-20}) $ & $I_{3}=($-$9.71\times 10^{-20}- $-$7.77 \times 10^{-20})\oplus (7.71\times 10^{-21}-9.71\times 10^{-20})$
\\
&$\rm{\Phi_{B2}}=$-$(14.8^{0}-9.1^{0})\oplus (9.5^{0}-14.9^{0})$&$\rm{\Phi_{B2}}=$-$(173.15^{0}-163.05^{0})\oplus (164.4^{0}-174.98^{0})$\\
\hline
 B3 & $I_{1}=
($-$1.08\times 10^{-11}- $-$5.19\times 10^{-12})\oplus (5.58\times 10^{-12}-1.09\times 10^{-11})$& $I_{1}=
($-$1.08\times 10^{-11}- $-$5.34\times 10^{-12})\oplus (5.45\times 10^{-13}-1.05\times 10^{-11})$  \\
 &$I_{2}=($-$2.19\times 10^{-8}- $-$2.32\times 10^{-10})\oplus (2.45\times 10^{-10}-2.33\times 10^{-8})$ 
&$I_{2}=($-$2.17\times 10^{-8}- $-$2.34\times 10^{-10})\oplus (2.65\times 10^{-10}-2.83\times 10^{-8})$  \\
&$I_{3}=
($-$2.75\times 10^{-17}- $-$1.38 \times 10^{-17})\oplus (1.40\times 10^{-17}-2.80\times 10^{-17}) $ & $I_{3}=($-$2.87\times 10^{-17}- $-$1.36 \times 10^{-17})\oplus (1.308\times 10^{-17}-2.88\times 10^{-17})$
\\
&$\rm{\Phi_{B3}}=$-$(17.59^{0}-10.24^{0})\oplus (10.55^{0}-17.50^{0})$&$\rm{\Phi_{B3}}=$-$(174.5^{0}-163.45^{0})\oplus (165.4^{0}-174.5^{0})$\\
\hline
B4 & $I_{1}=
($-$1.07\times 10^{-11}- $-$5.18\times 10^{-12})\oplus (5.81\times 10^{-12}-1.09\times 10^{-11})$& $I_{1}=
($-$1.09\times 10^{-11}- $-$5.33\times 10^{-12})\oplus (5.84\times 10^{-13}-1.05\times 10^{-12})$  \\
 &$I_{2}=($-$2.30\times 10^{-8}- $-$1.30\times 10^{-10})\oplus (1.31\times 10^{-10}-2.31\times 10^{-8})$ 
&$I_{2}=($-$4.40\times 10^{-8}- $-$2.35\times 10^{-10})\oplus (2.65\times 10^{-10}-4.82\times 10^{-8})$  \\
&$I_{3}=
($-$2.85\times 10^{-17}- $-$1.48 \times 10^{-17})\oplus (1.40\times 10^{-17}-2.89\times 10^{-17}) $ & $I_{3}=($-$2.97\times 10^{-17}- $-$1.46 \times 10^{-17})\oplus (1.48\times 10^{-17}-2.89\times 10^{-17})$
\\
&$\rm{\Phi_{B4}}=$-$(17.52^{0}-10.84^{0})\oplus (10.85^{0}-17.53^{0})$&$\rm{\Phi_{B4}}=$-$(175^{0}-165.45^{0})\oplus (166.4^{0}-174.9^{0})$\\
\hline

C & $I_{1}=
($-$9.58-9.52)\times 10^{-12}$& $I_{1}=
$-$(1.04\times 10^{-11}-5.26 \times 10^{-12})\oplus 5.14\times 10^{-12}-1.01 \times 10^{-12}  $  \\
 &$I_{2}=
($-$3.68-3.56) \times 10^{-7}$&$I_{2}=
$-$(7.74\times 10^{-8}-8.12 \times 10^{-10})\oplus 8.26\times 10^{-10}-7.14 \times 10^{-8}  $ \\
&$I_{3}=
($-$1.48-1.41)\times 10^{-21}$& $I_{3}=
$-$(1.24\times 10^{-22}-3.17 \times 10^{-24})\oplus 3.36\times 10^{-24}-1.24 \times 10^{-22}  $  \\
&$\rm{\Phi_{C}}=$-$(180^{0}-178.5^{0})\oplus (178.55^{0}-180^{0})$&$\rm{\Phi_{C}}=$-$(179.8^{0}-178.45^{0})\oplus (178.8^{0}-179.8^{0})$\\
\hline
\end{tabular}}
\caption{The predicted ranges of  $\rm{I_{1}, I_{2}  , I_{3}}$  and Jarlskog rephasing invariant parameter $J_{CP}$  for  the seven FGM cases of texture two zero has been presented. $\rm{I_{1}  , I_{2}  , I_{3}}$  are in $\rm{10^{54}(eV/c)^{12} , 10^{36}(eV/c)^{8}, (10^{54})^{2}(eV/c)^{18}}$,  respectively. The symbols - and $-$ denote mathematical minus and hyphen signs,  respectively. }\label{tab2}

\end{center}
\end{small}
\end{table}
\newpage
\section{Numerical analysis}
 As next step of the analysis, I shall discuss the numerical results of the analysis.
 For executing the analysis, I summarize the experimental information about various neutrino oscillation parameters. For both
normal mass ordering (NO) and inverted mass ordering (IO),  the latest experimental constraints on neutrino parameters at  3$\sigma$ confidence level (CL), following Ref. \cite{27}, are given below\\
\begin{small}
\begin{eqnarray*}
\rm{\delta m^{2}}[10^{-5}\rm{eV}^{2}]=7.05 - 8.14,\\
|\rm{\Delta m^{2}_{31}}|[10^{-3}\rm{eV}^{2}](NO)=2.41 - 2.60,\\
 |\rm{\Delta m^{2}_{31}}|[10^{-3}\rm{eV}^{2}](IO)=2.31 - 2.51, \\
  \theta_{12}=31.5^{\circ} - 38^{\circ},\\
  \theta_{23} (\rm{NO})=41.8^{\circ}- 50.7^{\circ},\\
  \theta_{23} (\rm{IO})=42.3^{\circ}-50.7^{\circ},\\
  \theta_{13} (\rm{NO})= 8.0^{\circ}- 8.9^{\circ},\\
  \theta_{13} (\rm{IO})= 8.1^{\circ}- 9.0^{\circ}.
  \end{eqnarray*}
\end{small}
 The inputs regarding charged leptons i.e.  electron, muon, tau neutrinos, are provided below
\begin{eqnarray*}
\rm{m_{e}}=0.000510998928 \rm{GeV},\\
\rm{m_{\mu}}=0.1056583715 \rm{GeV},\\
\rm{m_{\tau}}=1.77682 \rm{GeV}.
\end{eqnarray*}

After having discussed the analytical expressions for CP phases, it becomes imperative to calculate, $\rm{I_{1}, I_{2}}$ and $ \rm{I_{3}}$ , and its implications  for all the seven viable ansatz of texture two zero mass matrix for both  NO and IO. In Table \ref{tab2}, I have encapsulated the ranges of WB invariants  using the above input data. The multiplicative factor  in the caption arises  due to charged lepton masses $\rm{m_{e}}, \rm{m_{\mu}}, \rm{m_{\tau}}$, and difference in the factor is due to variable dependence on the terms pertaining to  $\rm{m_{e}}, \rm{m_{\mu}}, \rm{m_{\tau}}$. 
For calculating the invariants, I have adopted the  method of random number generation for generating the data points for input parameters, mass squared differences ($\rm{\delta m^{2}}$, $\rm{\Delta m^{2}_{31}}$),  neutrino mixing angles ($\theta_{12}, \theta_{23}, \theta_{13}$)   within 3$\sigma$ error of neutrino oscillation data, as given above. On the other hand, $\delta$ is given full variation from $0^{\circ}-360^{\circ}$ in the absence of any concrete experimental constraint. 
 
\begin{figure}[ht]
\begin{center}
\subfigure[]{\includegraphics[width=0.40\columnwidth]{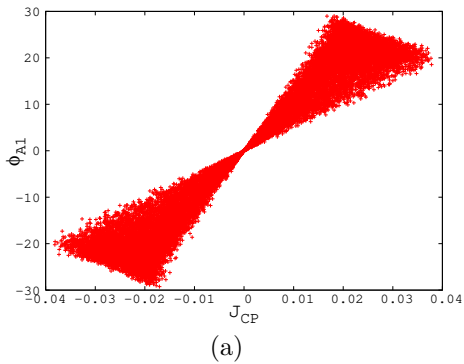}}
\caption{\label{fig1} Plot between Jarlskog rephasing invariant parameter $\rm{J_{CP}}$ and  reactor mixing angle  $\Phi_{A1}$ for NO at 3 $\sigma$ CL.   Angle $\Phi_{A1}$ is in degree. }
\end{center}
\end{figure}

 From the  discussion in section III, it is found that $\rm{J_{CP}}$ is the only experimentally feasible parameter. In this regard, I have provided the  analytical relations for $\rm{J_{CP}}$ for each viable ans\"atz. Hence it is convenient to calculate the allowed range of $\rm{J_{CP}}$ using the neutrino oscillation data.
As found in Ref. \cite{29}, ans\"atze \textbf{A1} and \textbf{A2} predict NO, while IO is ruled out at 3$\sigma$ CL. On the other hand, ans\"atze \textbf{B1, B2, B3, B4} and \textbf{C} predict both NO and IO at same CL. Using the analytical expression in Eq.(24), it is found that $\rm{J_{CP}}$  ranges from  $-$0.0377 to 0.0377, implying that  both CP violation and conservation are allowed [Fig{\ref{fig1}}. \\

Regarding the CP phase, the only phase which is directly measurable is relative phase of the elements of quartet of Q. For instance,  \\
\begin{equation*}
\rm{\Phi_{A1}\equiv argQ_{A1}=arg(m_{\mu\mu}m_{\tau\tau}(m_{e\tau}^{*})^{2} )},
\end{equation*}
corresponds to ans\"atz \textbf{A1}. Similarly, it is trivial to express the relative phase for remaining ans\"atze.
Using the neutrino oscillation data, it is found that  $\Phi_{\rm{A1}}=-28.9^{0}-28.5^{0}$, hence $\Phi_{\rm{A1}}$ seems to be unrestricted,  and hence  both the possibilities for CP  are open for ans\"atz \textbf{A1}.  Fig.\ref{fig1} reveals the strong correlation  between  the $\rm{J_{CP}}$ and $\Phi_{\rm{A1}}$ for  \textbf{A1}.
 From Table \ref{tab2}, the calculated range of $\rm{I_{1}}$ for these ansatz further affirms these predictions since $\rm{I_{1}}$ is directly proportional to $\rm{J_{CP}}$  \cite{14}. 
 
\begin{figure}[ht]
\begin{center}
\subfigure[]{\includegraphics[width=0.40\columnwidth]{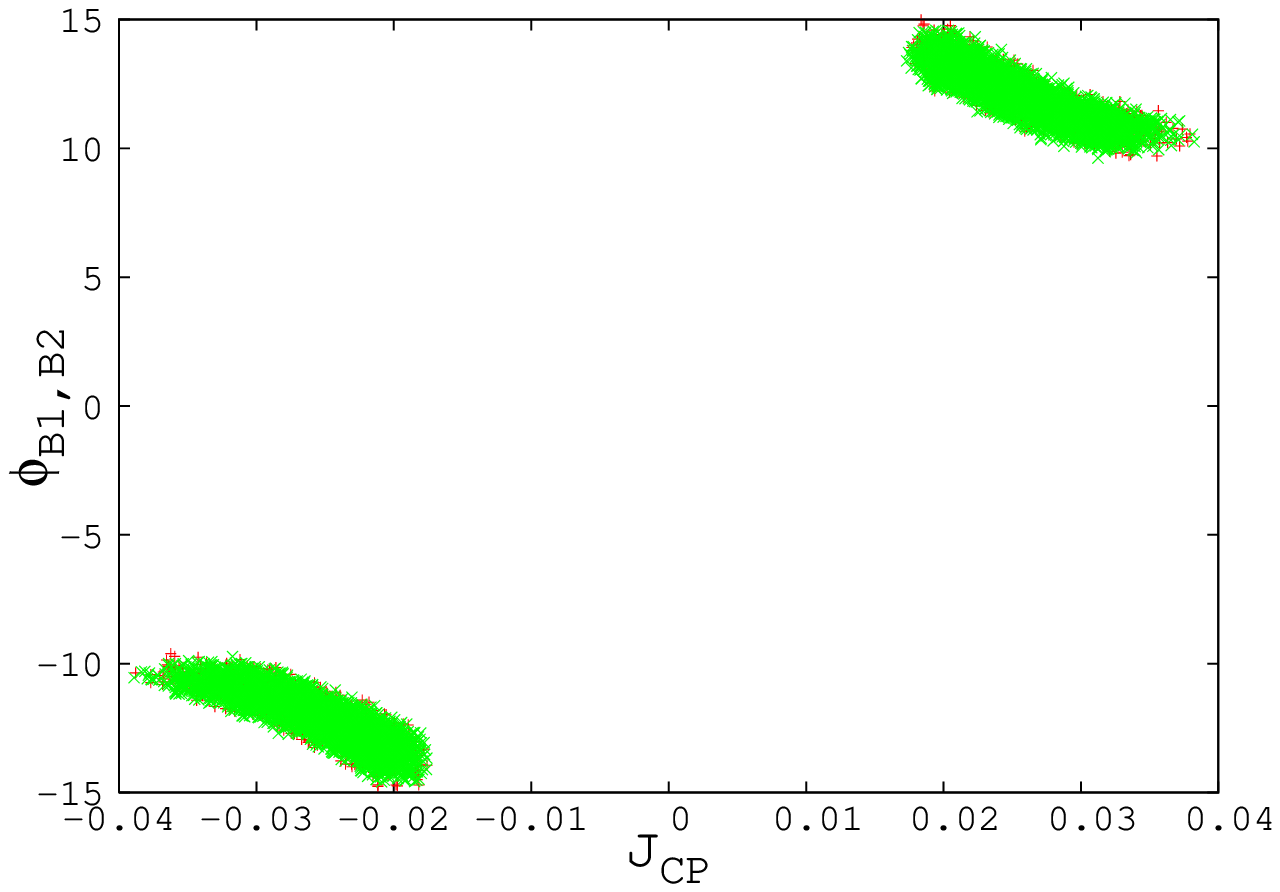}}
\subfigure[]{\includegraphics[width=0.40\columnwidth]{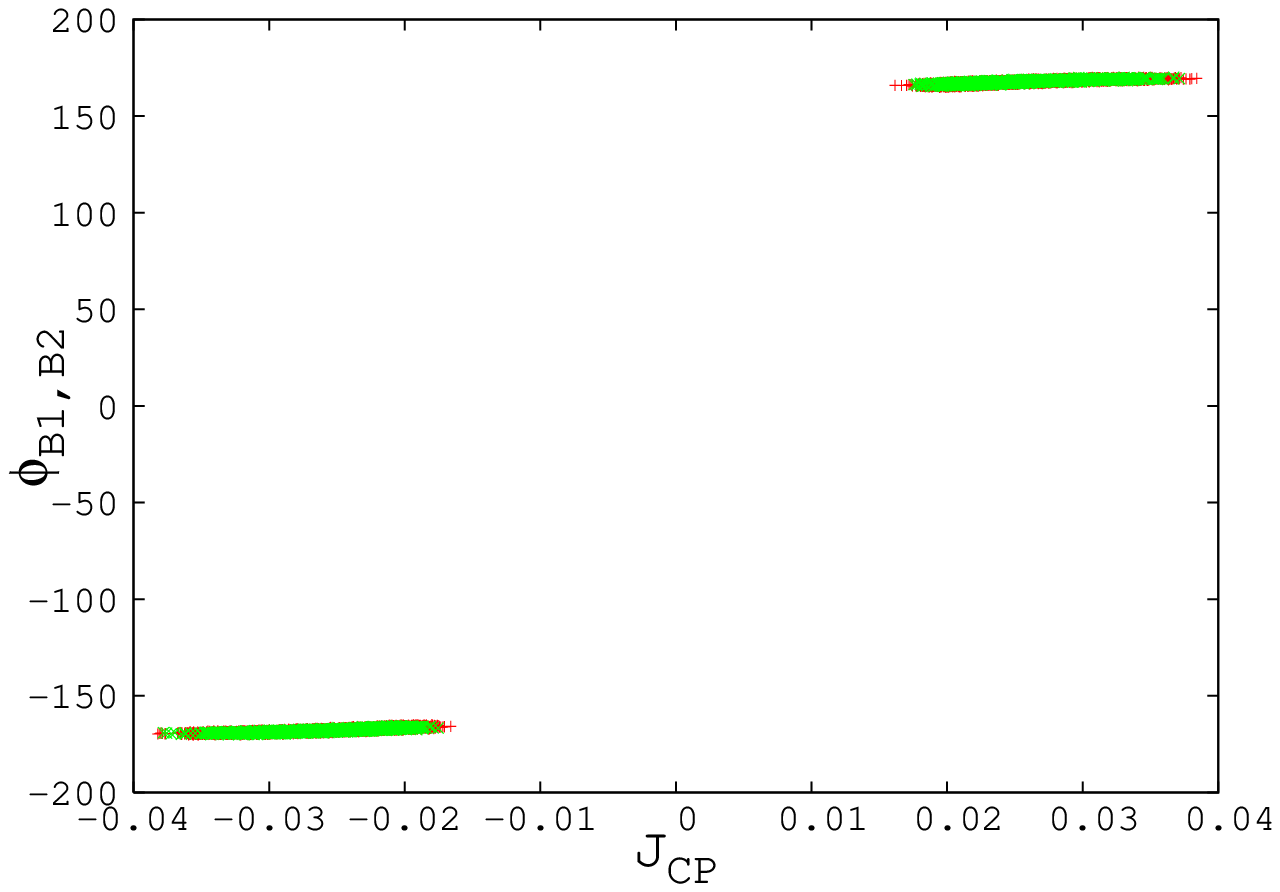}}\\
\caption{\label{fig2} Plot between Jarlskog rephasing invariant parameter $\rm{J_ {CP}}$ and  reactor mixing angle  $\Phi_{B1,B2}$ for  ansatze  \textbf{B1} and \textbf{B2}, (a) NO (b) IO. The red  and green color corresponds to \textbf{B1} and \textbf{B2}, respectively. Angle $\Phi_{B1,B2}$ is in degree. }
\end{center}
\end{figure}
\begin{figure}[ht]
\begin{center}
\subfigure[]{\includegraphics[width=0.40\columnwidth]{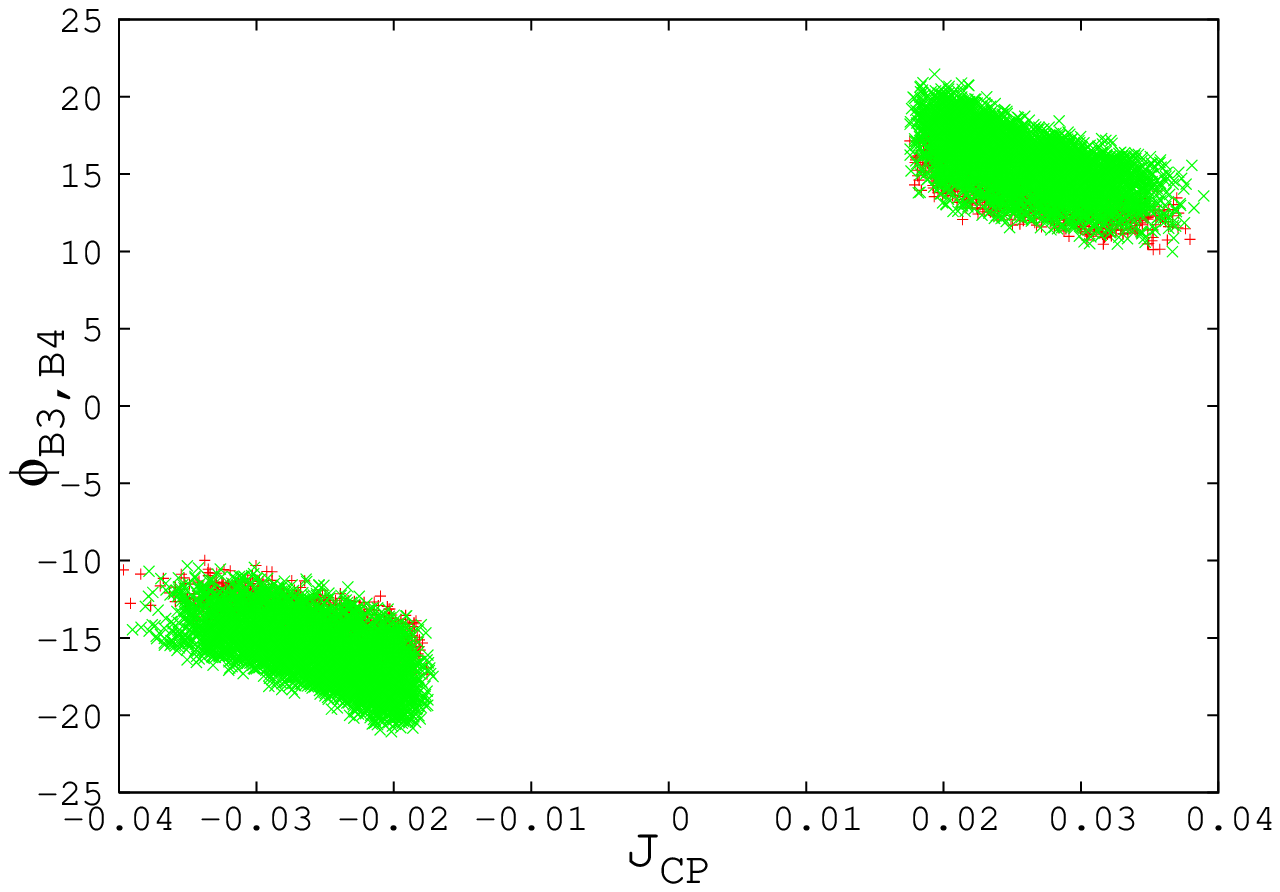}}
\subfigure[]{\includegraphics[width=0.40\columnwidth]{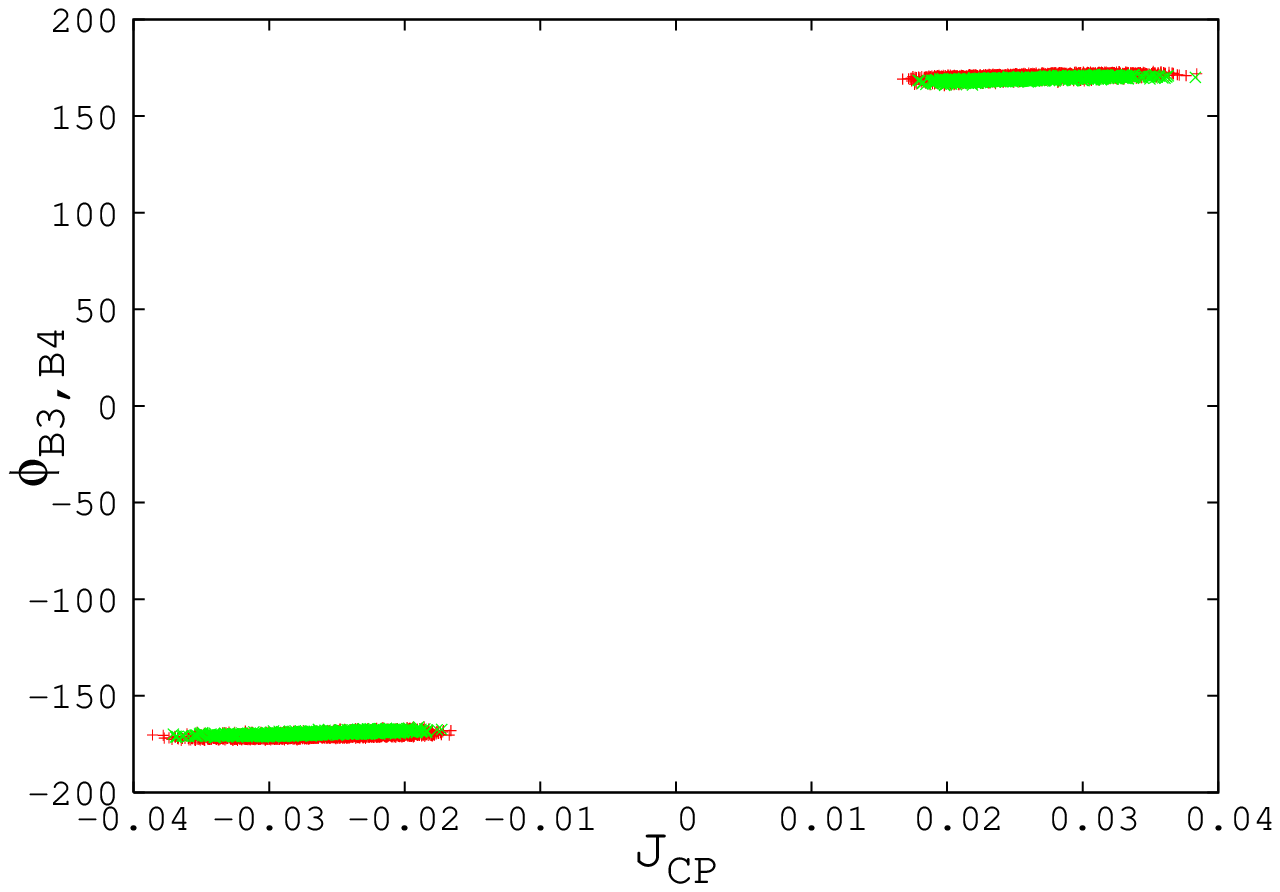}}\\
\caption{\label{fig3} Plot between Jarlskog rephasing invariant parameter $\rm{J_ {CP}}$ and  relative phase  angle  $\Phi_{\rm{B3,B4}}$ for  \textbf{B3} and \textbf{B4} (a) NO (b) IO. The red  and green color corresponds to \textbf{B1} and \textbf{B2}, respectively.  Angle $\Phi_{\rm{B3, B4}}$ is in degree. }
\end{center}
\end{figure}
\begin{figure}[ht]                                  
\begin{center}
\subfigure[]{\includegraphics[width=0.40\columnwidth]{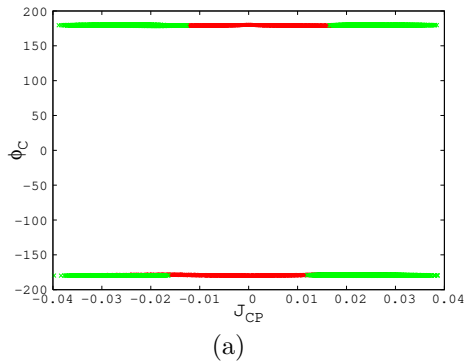}}
\caption{\label{fig4} Plot between Jarlskog rephasing invariant parameter $\rm{J_ {CP}}$ and  relative phase angle  $\Phi_{C}$ for  \textbf{C}. The red and green color corresponds to   NO  and IO, respectively. Angle $\Phi_{C}$ is in degree. }
\end{center}
\end{figure}

 For  ans\"atze belonging to \textbf{B},  I have complied the correlation plots for $\rm{J_{CP}}$ and $\Phi_{Bi}$ (i=1,2,3,4),  at 3$ \sigma$ CL [Figs. \ref{fig2},\ref{fig3} ]. As explicitly shown in Fig. \ref{fig2}, Fig. \ref{fig3} and Fig. \ref{fig4}, $\rm{J_{CP}}$ is non zero, implying that $\rm{I_{1}}$ is nonzero. From Table \ref{tab2}, it is apparent that $\rm{I_{1}, I_{2}, I_{3}}$ are found to be non-zero, implying that, CP is strictly violating for these ans\"atze in both lepton number conserving processes and violating processes. In addition, $\rm{J_{CP}}$ is nearly maximal,  which further implies that CP is maximally violated for these ansatz[Fig.\ref{fig2}, Fig.\ref{fig3}.
It must be remembered that  maximal CP violation implies $\delta\simeq 90^{0}$ (or $270^{0})$. 
In Ref. \cite{26}, using the $\chi^{2}$ analysis, the maximum allowed CP violation,   $\rm{J_{CP}^{max}}= 0.0329 \pm 0.0009 (\pm 0.0027)$ is determined at 1$\sigma$ (3$\sigma$) for both orderings. The preference of the present data for non-zero $\delta$ implies a best fit  $\rm{J_{CP}^{max}}= $ $-$0.032, which is favored over CP conservation at the $\sim 1.2 \sigma$ level. Therefore at present, our results are in tune with these results.\\
Regarding the relative phase, it is calculated as,  $\Phi_{\rm{B1}}=$-$(15^{0}-9.4^{0})\oplus (9.35^{0}-15^{0})$ for NO, and $\Phi_{\rm{B1}}=$-$(172.15^{0}-162.05^{0})\oplus (162.4^{0}-171.98^{0})$ for IO. Therefore, the possibility of CP invariance is ruled out, although the results are found  in close approximation to CP invariance condition.
Similarly, the  results for remaining ans\"atze of \textbf{B} are found to be same as \textbf{B1} and seems indistinguishable [table \ref{tab2}]. Apparently, the phases belonging to NO and IO are related as, $\Phi_{\rm{Bi}}$ (NO)$\simeq$ $180^{0}$- $\Phi_{\rm{Bi}}$(IO), where i=1, 2, 3, 4.\\
On comparing the  numerical results for WB invariants $\rm{I_{1,2,3}}$, it is apparent from the Table $\ref{tab2}$, 
$\rm{I_{1}}$ is almost indistinguishable for all the ans\"atz belonging to \textbf{B}.\\
 The phenomenological results for \textbf{B1 (B3)} and  \textbf{B2 (B4)} are almost identical and related to each other through $\mu-\tau$ permutation symmetry. For $\rm{B_{1}}(\rm{B_{2}})$, $\rm{I_{2}}=0$ and $\rm{I_{2} }\neq 0$ for $\rm{B_{3}}(\rm{B_{4}})$. 
On the other hand, $\rm{I_{3}}$ seems to  distinguish the \textbf{B1 (B2)} and \textbf{B3 (B4)}. The magnitude of $\rm{I_{3}}$ for B3 (B4) is greater than for B1(B2) by the order of $10^{3}$.\\  
 Therefore it is observed that $\rm{I_{2}}$ and $\rm{I_{3}}$, which are sensitive to Majorana type  CP violation,  appear  to distinguish the otherwise indistinguishable ansatze.  
  \\
   
   For ansatz \textbf{C},  it is found that for NO, both the possibilities of CP  are open, while for IO, CP conservation is not allowed if numerical range of $\rm{I_{1,2,3}}$ is taken into account [Table \ref{tab2}]. Therefore, it is possible to distinguish NO and IO for \textbf{C}. 
 On the other hand, $\Phi_{C}$ is found to be very close to CP invariance condition, $\Phi_{C}\simeq 180^{0}$ [Fig. \ref{fig4}]. More specifically, for NO, the parameter space of $\Phi_{C}$  allows CP invariance condition, while, for IO, it  marginally excludes the CP invariance condition. 
	 
\begin{figure}[ht]                                  
\begin{center}
\subfigure[]{\includegraphics[width=0.40\columnwidth]{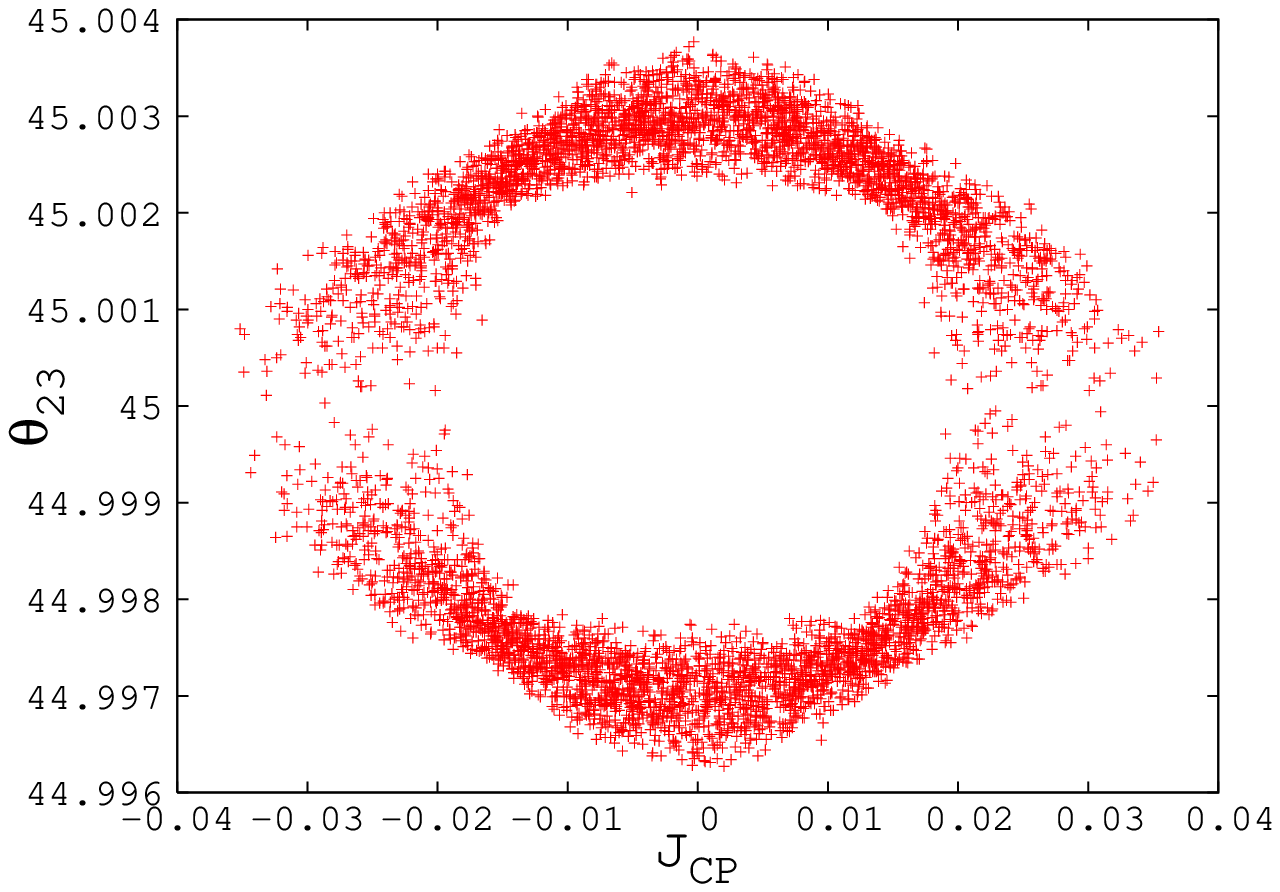}}
\subfigure[]{\includegraphics[width=0.40\columnwidth]{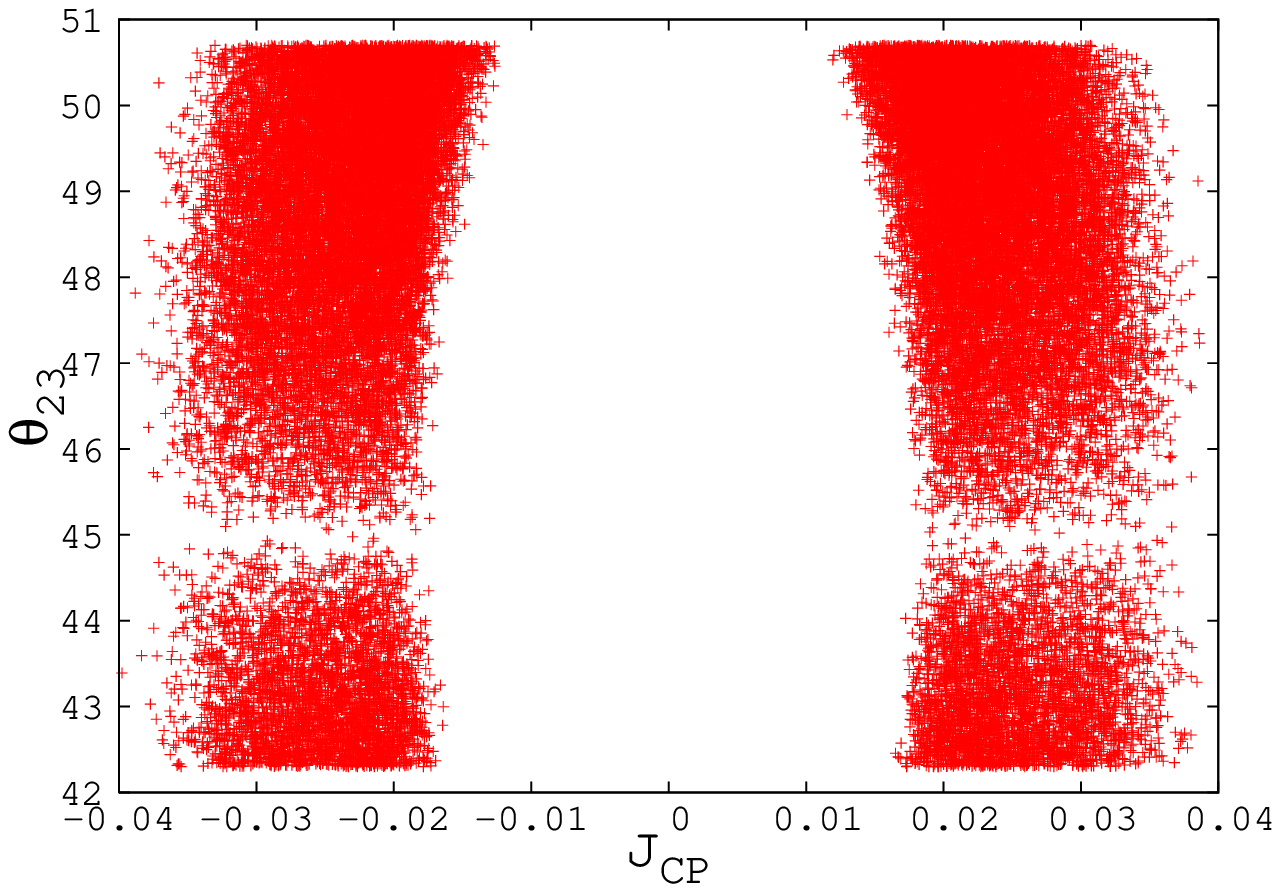}}
\caption{\label{fig5} Plot between Jarlskog rephasing invariant parameter $J_ {CP}$ and  atmospheric  mixing angle  $\theta_{23}$ for  \textbf{C}  (a) NO (b) IO. Angle $\theta_{23}$ is in degree. }
\end{center}
\end{figure}

Regarding the Jarlskog invariant parameter, $\rm{J_{CP}}$, it is found that $\rm{J_{CP}}$ =0 in NO,  which may be linked to the $\mu-\tau$ permutation symmetry as described in section II. The  reason can be understood, through the expression of s, given in subsection 3.3. From Fig.[\ref{fig5}(a)], it is apparent that $\theta_{23}\simeq 45^{0}$ for NO, which contributes to, s$\simeq O(10^{-6} - 10^{-7})$ after taking into account the experimental inputs of oscillation parameters. Therefore , $\rm{J_{CP} \propto (|m_{e\mu}|^{2}-|m_{e\tau}|^{2})}$ approaches to zero. \\
On the other hand, in case of IO, $\theta_{23}$ is deviated more from $45^{0}$, and also asymmetric around $45^{0}$[Fig. \ref{fig5}(b)], therefore a gap between the difference, $\rm{(|m_{e\mu}|^{2}-|m_{e\tau}|^{2})},$ is relatively more. This primarily explains why for NO,  $\rm{J_{CP}}$ =0 , and  for IO, $\rm{J_{CP}}\neq 0$.

\section{Summary and conclusions}

In  search of CP violation, I have re-investigated the CP-odd  WB invariants in terms of the neutrino mass matrix, and subsequently studied their implications for texture two zero neutrino mass matrix. The invariants pertaining to viable ansatz of texture two zero neutrino mass matrix, have common origin and linked  with the general form  of quartets of $\rm{m_{\nu}}$.   In an attempt to update the results of S. Dev \cite{19}, I have computed the CP phases $\rho, \sigma, \delta$ in terms of the neutrino oscillation parameters, while taking into account  the CP invariance condition, for each ans\"atz. As a next step, I have performed the numerical analysis, and  calculated the ranges of WB invariants $\rm{I_{1}, I_{2}, I_{3}}$ for viable texture two ansatze using neutrino oscillation data. It has been found that 
ans\"atze \textbf{A1, A2, C} appear to predict both the possibilities viz. CP violation as well as  CP conservation, while ansatze \textbf{B1, B2, B3, B4 } violate CP symmetry. The analysis further comment on the distinguishabilty of the ans\"atze belonging to \textbf{B} using invariants $\rm{I_{1}, I_{2}, I_{3}}$, which are other wise not distinguishable.  The invariant $\rm{I_{1}}$, which is sensitive to lepton number conserving process (LNC), can not distinguish ans\"atze \textbf{B1, B2, B3, B4 }, while $\rm{I_{2}}$ and $\rm{I_{3}}$ pertaining to lepton number violationg process (LNV)
 can distinguish \textbf{B1, B2, B3, B4 }. The analytical expressions for 
 Jaralskog rephasing invariant  have been computed for each ansatz. Using the oscillation data, it is found that for ans\"atze  belong to \textbf{B}, the parametric space of $\rm{J_{CP}}$ is restricted near  maximal CP violation, hence remain  compatible with the latest hints provided in neutrino oscillation data. For ans\"atz \textbf{C}, $\rm{J_{CP}}$ plays a vital role   to distinguish NO and IO.
 At present, these indications predict  CP violation and signal towards the possible CP violations in both LNC and LNV process, respectively, however only the future double beta decay experiments (LNV process) and neutrino oscillations (LNC processes) as well as the cosmological data could throw some light on these predictions.\\
\\
\section*{Data availabilty}

The data used to support the findings of this study is included within this article.

\section*{Conflicts of Interest}
The authors declare that there are no conflicts of interest regarding the publication of this paper.

\section*{Acknowledgment}
The author would like to thank the Principal of M. N. S. Government College,  Bhiwani, Haryana for providing necessary facilities to work. \\


\begin{thebibliography}{26}


\bibitem{1} S. L. Glashow, Nucl. Phys. 22, 597 (1961).

\bibitem{2} S. Weinberg, Phys. Rev. Lett.19, 1264 (1967).

\bibitem{3} A. Salam, Proc. of the 8th Noble Symposium on Elementary Particle Theory, Relativistic Groups and Analyticity, edited by N.Svartholm (1969).

\bibitem{4}  H. Fritzsch, Gell-Mann and H. Leutwyler, Phys. Lett. B 4, 365 (1973).

\bibitem{5}  For excellent reviews on the Standard Model see, J. F. Dooghue, E. Golowich and     B. R.Holstein, Dynamics of the Standard Model, Cambridge University Press, 1992.

\bibitem{6}  N. Cabbibo, Phys. Rev. Lett. 10,  531 (1963); M. Kobayashi and K. Maskawa,     Prog. Theor. Phys. B49, 652 (1973).

\bibitem{7} M. Tanabashi et al. (Particle Data Group) Phys. Rev. D 98 030001 (2018), updated results available at http://pdg.lbl.gov/.

\bibitem{8}  J. Charles et al., CKMfitter Group, updated result available at    http://www.ckmfitter.in2p3.fr/.

\bibitem{9} A. J . Bevan et al. [ Utfit Collaboration], JHEP 03, 123 (2014), updated results available at http://www.utfit.org/.

\bibitem{10} Y. Amhis et al., Heavy Flavor Averaging Group (HFAG), hep-ex/1207.1158 v2(2013),  updated results available at http: www.slac.stanford.edu/xorg/hfag/.

\bibitem{11} R. Davis, Prog. Part. Nucl. Phys. 32, 13 (1994); B.T. Cleveland et al., Astrophys. J. 496, 505 (1998); S. Fukuda et al.,SuperKamiokande Collaboration, Phys. Rev. Lett. B 539, 179 (2002); Q. R. Ahmad et al., SNO  Collaboration Phys. Rev. Lett. 89, 011301(2002); S. N Ahmad et al., Phys. Rev. Lett. 92, 181301 (2004).

\bibitem{12} Y. Fukuda et al. SuperKamiokande Collaboration, Phys. Rev. Lett. 81, 1562 (1998); A. Surdo, MACRO Collaboration, Nucl. Phys. Proc. Suppl. 110, 342 (2002); M. Sanchez, Soudan  Collaboration, Phys, .Rev D68, 113004 (2003).

\bibitem{13}  K. Eguchi et al., KamLAND Collaboration, Phys. Rev Lett. 90, 021802 (2003); Phys. Rev. Lett. 94, 081801 (2005).

\bibitem{14}  M. H. Ahn et al., K2K Collaboration, Phy. Rev. Lett. 90, 041801 (2003).
\bibitem{15} H. Fritzsch, Z. Z. Xing, Phys. Lett. B 517 (2001) 363-368, arXiv: hepph/0103242
\bibitem{16}   B. Pontecorvo , Zh. Eksp. Teor. Fiz. (JETP) 33, 549 (1957); ibid. 34, 257 (1958); ibid. 53, 1717(1967); Z. Maki, M. Nakagawa, S. Sakata, Prog. Theor. Phys. 28, 870(1962).
\bibitem{17} Y. Farzan and A. Y. Smirnov, Phys. Rev. D 65 (2002) 113001 [arXiv:hep-ph/0201105].
\bibitem{18} J. A. Aguilar-Saavedra and G. C. Branco, Phys. Rev. D 62 (2000) 096009
[arXiv:hep-ph/0007025]; J. Sato, Nucl. Instrum. Meth. A 472 (2000) 434
[arXiv:hep-ph/0008056].
\bibitem{19}  S. Dev, Sanjeev Kumar, Surender Verma, Phys. Rev. D 79, 033011, (2009).
\bibitem{20}  Utpal Sarkar and Santosh K. Singh, Nucl. Phys. B 771, 28-39 (2006).
\bibitem{21}   S. Dev, Shivani Gupta, R. R.  Gautam, J. Phys. G37:125003, 2010.
\bibitem{22}  Y. Abe et. al. [Double Chooz collaboration], Phys. Rev. Lett. 108, 131801 (2012).

\bibitem{23}  F. P. Ann et. al. [Daya Bay collaboration], Phys. Rev. Lett. 108, 171803 (2012).

\bibitem{24}  Soo-Bong Kim, [RENO collaboration], Phys. Rev. Lett. 108, 191802 (2012).

\bibitem{25}  K. Abe et. al. [T2K collaboration], Phys. Rev. Lett. 107, 041801 (2011).
\bibitem{26} I. Esteban et al, JHEP 01 (2019) 106, arXiv: 1811.05487v1 [hep-ph].
\bibitem{27} P. F. de Salas, D. V. Forero, C. A. Ternes, M. Tortola, J. W. F. Valle, Phys. Lett. B 782 (2018) 633, arXiv: 1708.01186 [hep-ph].
\bibitem{28}Z. Z. Xing and S. Zhou, “Neutrinos in particle physics, astronomy and cosmology,” SpringerVerlag, Berlin Heidelberg (2011).
\bibitem{29} H. Fritzsch, Z. Z. Xing, S. Zhou, JHEP 1109, 083 (2011).
\bibitem{30} C. Jarlskog, Phys. Rev. Lett. 55, 1039 (1985).
\bibitem{31} G. C. Branco, L. Lavoura and M. N. Rebelo, Phys. Lett. B 180 (1986) 264.
\bibitem{32} G. C. Branco, M. N. Rabelo, J. I. Silva-Marcos Phys. Rev. Lett. 82 (1999) 683.
\bibitem{33} Y. Farzan, and A. Y. Smirnov, JHEP 01, 059 (2007), arXiv: hep-ph/061037v3.
\bibitem{34} B. Yu, S. Zhou, Phys. Lett. B 800 (2020) 135085, arXiv: 1908.09306 [hep-ph].
\bibitem{35} R. Samanta, M. Chakraborty, A. Ghosal, Nucl. Phys. B,  904 (2016), 86-105, arXiv: 1502.06508v2 [hep-ph].



%
%
%\bibitem{1}  Y. Abe et. al. [Double Chooz collaboration], Phys. Rev. Lett. 108, 131801 (2012).
%
%\bibitem{2}  F. P. Ann et. al. [Daya Bay collaboration], Phys. Rev. Lett. 108, 171803 (2012).
%
%\bibitem{3}  Soo-Bong Kim, [RENO collaboration], Phys. Rev. Lett. 108, 191802 (2012).
%
%\bibitem{4}  K. Abe et. al. [T2K collaboration], Phys. Rev. Lett. 107, 041801 (2011).
%\bibitem{5}  P. H. Frampton, S. L. Glashow, D. Marfatia, Phys. Lett. B 536, 79 (2002); Z. Z. Xing, 
%       Phys.   Lett. B 530, 159 (2002); B. R. Desai, D. P. Roy, A. R. Vaucher, Mod. Phys. Lett. 
%       A    18, 1355 (2003); Z. Z. Xing, Int. J. Mod. Phys. A 19, 1 (2004); A. Merle, 
%       W. Rodejohann, Phys. Rev. D 73, 073012 (2006); S. Dev, S. Kumar, S. Verma, S. Gupta, 
%       R.  R. Gautam, Phys. Rev. D 81, 053010 (2010) and references therein; H. Fritzsch, 
%       Z. Z. Xing, S. Zhou, JHEP 1109, 083 (2011); X. W. Liu, S. Zhou, Int. J. Mod. Phys. 
%       A 28,1350040 (2013); D. Meloni, A. Meroni, E. Peinado, Phys. Rev. D 89, 053009 (2014).
%
%\bibitem{6}  M. Fukugita, M. Tanimoto, T. Yanagida, Prog. Theor. Phys. 89, 263 (1993);ibid. 
%      Phys. Lett. B 562, 273 (2003); M. Fukugita, et al., Phys. Lett. B 716, 294
%      (2012).
%
%\bibitem{7} M. Gupta, G. Ahuja, Int. J. Mod. Phys. A 26 (2011) 2973; ibid. Int. J. Mod.
%      Phys. A 27 (2012) 1230033 and references therein.
%
%\bibitem{8}  H. Fritzsch, Z. Z. Xing, S. Zhou, JHEP 1109, 083 (2011).
%
%\bibitem{9} J. Y. Liu and S. Zhou, Phys. Rev. D87, 093010 (2013).
%
%\bibitem{10} S. Kaneko, H. Sawanaka and M. Tanimoto, J. High Energy Phys. 08 (2005) 073;
%
%\bibitem{11} L. Lavoura, Phys. Lett. B 609, 317 (2005), hep-ph/0411232; E. I. Lashin and N. Chamoun,
%Phys. Rev. D 78, 073002 (2008), arXiv: 0708.2423 [hep-ph];E. I. Lashin, N. Chamoun, Phys.
%Rev. D 80, 093004 (2009), arXiv:0909.2669 [hep-ph]; S. Dev, S. Verma, S. Gupta and R.
%R. Gautam, Phys. Rev. D 81,053010 (2010), arXiv: 1003.1006 [hep-ph]; S. Dev, S. Gupta
%and R. R. Gautam, Mod. Phys. Lett. A 26, 501-514 (2011), arXiv: 1011.5587 [hep-ph]; T.
%Araki, J. Heeck and J. Kubo, JHEP 1207, 083 (2012), arXiv: 1203.4951 [hep-ph].
%
%\bibitem{12} Madan Singh, Gulsheen Ahuja, Manmohan Gupta,  Prog. Theor. Exp. Phys. (PTEP) 
%         2016 (12): 123B 08, arXiv: 1603.08083 [hep-ph].

%\bibitem{13}   P. H. Frampton, S. L. Glashow, D. Marfatia, Phys. Lett. B 536, 79 (2002).
%
%
%
%
%\bibitem{17} Gustavo C. Branco, M. N. Rabelo, J. I. Silva-Marcos, Phys. Lett. B 633 (2006) 345-354.
%\bibitem{18} Gustavo C. Branco, M. N. Rabelo, New J. Phys. 7 86 (2005).
%\bibitem{19} G. C. Branco and M. N. Rebelo, Nucl. Phys. B 278, 738 (1986); Herbi K. Dreiner, 
%        Jong  SooKim, Oleg Lebedev, Marc Thormeier, hep-ph/0703074.
%\bibitem{20} C. Jarlskog, Phys. Rev. Lett. 55, 1039 (1985).
%\bibitem{21} G. C. Branco, L. Lavoura and M. N. Rebelo, Phys. Lett. B 180 (1986) 264.
%\bibitem{22} Gustavo C. Branco, M. N. Rabelo, J. I. Silva-Marcos Phys. Rev. Lett. 82 (1999) 683; 
%        Herbi K. Dreiner, Jong Soo Kim, Oleg Lebedev and Marc Thormeier, Phys. Rev. D 76
%  ,     015006 (2007) , hep-ph/0703074.
%\bibitem{23} I. Esteban, et al., JHEP 01, 087 (2017),  arXiv: 1611.01514 [hep-ph]; P. F. de Salas, D. V. Forero, C. A. Ternes, M. Tortola, J. W. F. Valle, arXiv:1708.01186 [hep-ph]..
%
%\bibitem{24}   M. C. Gonzalez-Garcia, M. Maltoni, J. Salvado and T. Schwetz,  arXiv:1409.5439 [hep-ph].
%
%\bibitem{25} F. T. Avignone III, S. R. Elliott, J. Engel, Rev. Mod. Phys. 80, 481 (2008), arXiv:0708.1033 [nucl-ex]; J. J. Gomez-Cadenas, J. Martin-Albo, M. Mezzetto, F. Monrabal, M. Sorel, arXiv:1109.5515 [hep-ex];S. M. Bilenky, C. Giunti, Mod. Phys. Lett. A 27, 1230015, arXiv:1203.5250 [hep-ph].
%
%\bibitem{26} W. Rodejohann, Int. J. Mod. Phys. E, 20, 1833 (2011), arXiv:1106.1334 [hep-ph].
%\bibitem{27} Manmohan Gupta,
%Gulsheen Ahuja,
%Madan Singh, Dheeraj Shukla (2018), DOI: 10.1142/9789813231801$\_$002, Conference: C17-02-06.3, p.229-242, https://www.worldscientific.com/doi/abs/10.1142/9789813231801$\_$0020.
%
%

\end{thebibliography}
\end{document}